\documentclass[journal,twocolumn]{IEEEtran}

\ifCLASSINFOpdf

\else

\fi

\usepackage{graphicx}
\usepackage{amsmath}
\usepackage{mathrsfs}
\usepackage[noend]{algpseudocode}
\usepackage{algorithmicx,algorithm}
\usepackage{amsthm}
\usepackage{amsfonts}
\usepackage{subfigure} 
\usepackage{color}
\usepackage{cite}
\usepackage{setspace}
\newtheorem{proposition}{Proposition}

\usepackage{amssymb}

\usepackage{float} 

\usepackage{xcolor}
\usepackage{xpatch}\usepackage{xcolor}
\usepackage{xpatch}
\makeatletter
\def\changeBibColor#1{%
	\in@{#1}{}
	\ifin@\color{blue}\else\normalcolor\fi	
}
\xpatchcmd\@bibitem
{\item}
{\changeBibColor{#1}\item}
{}{\fail}
\xpatchcmd\@lbibitem
{\item}
{\changeBibColor{#2}\item}
{}{\fail}
\makeatother

\begin{document}

\title{Windowing Optimization for Fingerprint-Spectrum-Based Passive Sensing in Perceptive Mobile Networks} 

\author{Xiao-Yang~Wang,~Shaoshi~Yang,~\IEEEmembership{Senior Member,~IEEE}, Hou-Yu~Zhai,~Christos~Masouros,~\IEEEmembership{Fellow,~IEEE},~J.~Andrew~Zhang,~\IEEEmembership{Senior Member,~IEEE}
\thanks{
	Part of this work has been accepted by the 2024 IEEE Global Communications Conference (GLOBECOM) \cite{wxyglobecom}.
	
	This work was supported in part by the Beijing Municipal Natural Science Foundation (No. L242013 and No. Z220004), in part by the National Key R\&D Program of China (No. 2023YFB2904803), in part by the Guangdong Major Project of Basic and Applied Basic Research (No.  2023B0303000001), in part by the Fundamental Research Funds for the Central Universities (No. 2023ZCJH02), in part by the Smart Networks and Services Joint Undertaking (SNS JU) under the European Union’s Horizon Europe Research and Innovation Programme (No. 101139176), and in part by BUPT Excellent Ph.D. Students Foundation (No. CX2023238). \textit{(Corresponding author: Shaoshi Yang.)}
	
	Xiao-Yang~Wang, Shaoshi~Yang and Hou-Yu Zhai are with the School of Information and Communication Engineering, Beijing University of Posts and Telecommnuications, the Key Laboratory of Universal Wireless Communications, Ministry of Education, and the Key Laboratory of Mathematics and Information Networks, Ministry of Education, Beijing 100876,	China. Xiao-Yang~Wang is also with the Department of Electronic and Electrical Engineering, University College London, London WC1E 7JE, UK (E-mail:\{wangxy\_028, 2hy, shaoshi.yang\}@bupt.edu.cn).}
\thanks{Christos~Masouros is with the Department of Electronic and Electrical Engineering,  University College London, London WC1E 7JE, UK (E-mail: c.masouros@ucl.ac.uk).}
\thanks{J.~Andrew~Zhang is with the Global Big Data Technologies Centre, University of Technology Sydney, Sydney NSW 2007, Australia (E-mail: andrew.zhang@uts.edu.au).}
}

\markboth{IEEE Transactions on Communications, accepted, Aug. 2024}%
{Shell \MakeLowercase{\textit{et al.}}: Bare Demo of IEEEtran.cls for IEEE Journals}

\maketitle

\begin{abstract}
Perceptive mobile networks (PMN) have been widely recognized as a pivotal pillar for the sixth generation (6G) mobile communication systems. However, the asynchronicity between transmitters and receivers results in velocity and range ambiguity, which seriously degrades the sensing performance. To mitigate the ambiguity, carrier frequency offset (CFO) and time offset (TO) synchronizations have been studied in the literature. However, their performance can be significantly affected by the specific choice of the window functions harnessed. Hence, we set out 
to find superior window functions capable of improving the performance of CFO and TO estimation algorithms. We firstly derive a near-optimal window, and the theoretical synchronization mean square error (MSE) when utilizing this window. However, since this window is not practically achievable, we then test a practical ``window function" by utilizing the multiple signal classification (MUSIC) algorithm, which may lead to excellent synchronization performance.
\end{abstract}


\IEEEpeerreviewmaketitle

\section{Introduction}
\IEEEPARstart{W}{ith} the explosive growth of wireless tele-traffic, cellular networks are resorting to ultra-broadband communications technologies in the millimeter wave (mmWave) or terahertz (THz) bands \cite{8373698,6998944}, which are naturally suited also for radar applications. With the assistance of mmWave and THz frequencies, high-precision sensing \cite{9858656,10226306}, such as range and velocity estimation, human activity sensing \cite{9941043,9724223} and imaging can be performed in mobile networks \cite{9737357}, resulting in the perceptive mobile network (PMN) concept \cite{rahman2019framework,zhang2021enabling,zhang2020perceptive, 9921271}. By exploiting the sensing capability, the PMN allows mutual benefits between the sensing and communication modules, such as an innovative framework that mutually enhances estimation accuracy and beamforming gain \cite{10226306} with power allocation. 
PMN relies on a pair of distinct sensing types: passive and active sensing \cite{zhang2020perceptive,article}. In passive sensing, the remote radio unit  (RRU) estimates the desired parameters by exploiting the reflected signals of the user equipments (UEs)  \cite{tong2021joint} or other RRUs \cite{zhang2020perceptive}. In active sensing mode, the RRU or UE works by processing the echo signals transmitted by itself \cite{10373185,liu2020joint,10086626,10061453}. Nevertheless, high-performance active  sensing requires practical full-duplex technology, which is not mature enough to be implemented efficiently at the time of writing \cite{liu2020joint,zhang2021overview}. Moreover, sensing in time-division duplex (TDD), such as \cite{10086626}, or frequency-division duplex (FDD), such as \cite{liu2020joint,9724258}, constrains the achievable communications rate \cite{zhang2021overview}. As a result, passive sensing offers a competitive alternative. 

However, challenges arise when aiming for high-precision passive sensing. The geographically separated transceivers of passive sensing systems inevitably use different oscillators, which leads to carrier frequency offset (CFO) and time offset (TO) between the transmitters and receivers \cite{4287203,6760595,10091198}. Furthermore, the CFO and TO will result in nonnegligible velocity and range sensing ambiguity, which severely degrades the sensing performance \cite{ni2021uplink}. To mitigate the ambiguity, both time and carrier frequency synchronization should be implemented \cite{10049817}. Nevertheless, due to the systematic difference between sensing in PMN \cite{4287203} and in bi-static radar \cite{1986IPCRS}, synchronization methods for traditional bi-static radar cannot be applied in PMN. Therefore, synchronization algorithms for passive sensing need to be carefully re-designed.

In passive sensing PMN systems, a limited number of synchronization methods have been reported \cite{zhang2021enabling,9848428,10418473}. The first method designed for passive sensing is the cross-antenna cross-correlation (CACC) algorithm \cite{IndoTrack}, which exploits the cross-correlation between signals received at different antennas to extract TO and CFO. However, the cross-correlation doubles the number of parameters to be estimated, which results in excessive computational complexity. To reduce the complexity, the mirrored multiple signal classification (MUSIC) algorithm was proposed in \cite{ni2021uplink}. This concept halves the number of estimated parameters, thus decreasing the overall computational complexity. However, these two algorithms are unsuitable for non-line-of-sight (NLOS) or single-antenna scenarios.
Another synchronization algorithm, namely FarSense \cite{FarSense}, uses the ratio of the channel state information (CSI) measurements corresponding to different antennas to extract the CFO and the up-to-date value of TO during a period in single-target scenarios. However, this algorithm cannot be applied to multi-target scenarios. To address this issue, the authors extended, FarSense to MultiSense \cite{zeng2020multisense}, where a filter is designed to separate signals reflected by different targets, enabling the estimation of the multi-pair values of TO and CFO. This modification allows MultiSense to operate in multi-target scenarios. Additionally, leveraging properties of the Mobius transform, the authors of \cite{9904500} propose a family of CSI-ratio-based frequency synchronization schemes, which can only mitigate CFO, in scenarios where a single target exists and its velocity is low.  To further characterize the impact of this kind of synchronization method on Doppler sensing, \cite{hu2024performance} derives the closed-from Cramér–Rao lower bound (CRLB) for Doppler sensing under CSI-ratio-based synchronization. However, the CSI-ratio based schemes in these works cannot support range estimation. Additionally, it is important to note that the aforementioned synchronization studies are conducted for multi-carrier passive sensing PMN based on existing mobile communication systems. Meanwhile, the authors of \cite{tagliaferri2023cooperative} propose a time and frequency synchronization method for cooperative multi-static imaging at a single frequency.

\begin{table*}[tbp]	\newcommand{\tabincell}[2]{\begin{tabular}{@{}#1@{}}#2\end{tabular}}  
	\small
	\centering
	\caption{Comparison between state-of-the-art CFO and/or TO estimation methods and our scheme for asynchronous sensing in PMN}
	\scalebox{1}{\begin{tabular}{c|c|c|c|c|c}  	
			\hline	
			\textbf{Methods}  & \tabincell{c}{\textbf{Propagation} \\ \textbf{Environment}} & \textbf{Number of Targets} & \tabincell{c}{\textbf{Sensing Functions} \\ \textbf{ Supported}} &\tabincell{c}{\textbf{Number of}\\ \textbf{Antennas}} &\tabincell{c}{\textbf{Number of}\\ \textbf{Subcarriers}}\\ \hline
			\tabincell{c}{CACC \cite{IndoTrack}, \\ mirrored-MUSIC \cite{ni2021uplink}} & LOS & {Multi}-target & velocity, range & Multi-antenna & Multi-subcarrier\\ \hline
			FarSense \cite{FarSense} & LOS / NLOS & Single-target&velocity & Multi-antenna & Multi-subcarrier\\ \hline
			MultiSense \cite{zeng2020multisense}& LOS / NLOS & Multi-target & velocity & Multi-antenna & Multi-subcarrier\\ \hline
			Schemes in \cite{9904500}& LOS / NLOS & Single-target & velocity & Multi-antenna & Multi-subcarrier\\ \hline
			Scheme in \cite{10207823}&LOS&Multi-target&velocity&Multi-antenna&Single-subcarrier\\
			\hline
			Scheme in \cite{tagliaferri2023cooperative}&LOS / NLOS&Multi-target&\tabincell{c}{cooperative\\ imaging}&\tabincell{c}{Single-antenna,\\ Multi-antenna}&Single-subcarrier\\
			\hline
			JUMP \cite{10443836} & LOS / NLOS & Multi-target & velocity, range & \tabincell{c}{Multi-antenna} & Multi-subcarrier\\
			\hline
			\tabincell{c}{Our scheme,\\ SHARP \cite{9804861}} & LOS / NLOS & Multi-target & velocity, range & \tabincell{c}{Single-antenna,\\ Multi-antenna} & Multi-subcarrier\\ \hline
	\end{tabular}} 
	\label{table2}
\end{table*}

The general framework of the aforementioned  synchronization schemes for multi-subcarrier PMNs involves extracting the TO and CFO by processing different signals received from different antennas with the aid of either cross-correlation methods, such as \cite{IndoTrack,ni2021uplink, 10616023}, or simple division operations, such as \cite{FarSense,zeng2020multisense}. However, this research framework \textit{restricts applications to multi-antenna scenarios}. In response to this limitation, a framework achieving synchronization by processing signals from different paths is explored \cite{10207823,10443836,9804861}. In principle, this framework does not require multiple receiving antennas. However, in \cite{10207823}, Doppler offset estimation at a single frequency  is implemented by separating each dynamic and static path with the estimated angle of arrival (AOA), still necessitating a multi-antenna receiver. In \cite{10443836}, although TO estimation can be achieved in single-antenna systems by exploiting the slow-changing delay profile of the channel impulse response, CFO estimation relies on high-resolution angle of departure (AOD), which cannot be obtained in single-antenna systems. Moreover, the authors of \cite{9804861} separate signals from multiple paths relying on a prior phase offset dictionary, which strongly depends on time delay, packet detection delay and carrier frequency offset. However, this work achieves synchronization by solving high-dimensional compressed sensing optimization problems with very high computational complexity. Overall, implementing these schemes in single-antenna PMNs is also challenging.

To further address synchronization in single-antenna systems, the cross-multipath cross-correlation (CMCC) synchronization algorithm in \cite{wxy} explored a novel framework with several key insights: 1) the delay-Doppler spectrum of the static signal component, namely the \textit{fingerprint spectrum} \cite{wxy}, is uniquely identified by location distribution of static objects reflecting signals; 2) the fingerprint spectrum will cyclically shift, when the CFO and TO drift. As a benefit of these insights, CMCC achieves high-precision CFO and TO estimation in LOS/NLOS and single-target/multi-target scenarios\footnote{Since the synchronization is implemented in the delay-Doppler domain, parameter estimation algorithms designed for orthogonal time frequency space (OTFS) modulation \cite{10147252}, such as the fractional delay and Doppler estimation \cite{10118873} and the compressed-sensing based channel estimation \cite{10472131}, can be potentially used to further improve the synchronization performance.}. For ease of comparison, we summarize the application scenarios for the aforementioned algorithms and CMCC in Table \ref{table2}. Concretely, in CMCC, the CFO and TO are acquired by locating the peak of the cross-correlation sequence of the fingerprint spectra at different instants of time. Since the fingerprint spectra are composed by 
the Fourier transform of the window functions utilized, the window function adopted naturally impacts the estimation performance \cite{wxy}. However, the existing contributions employ standard window functions and there is an open research horizon for re-designing the windows to further enhance the synchronization performance.

To improve the synchronization performance of CMCC, we conduct theoretical analysis to acquire superior window functions. 
In a nutshell, this research can be divided into two main parts. The first part focuses on designing an asymptotically optimal window for minimizing the synchronization mean square error (MSE), and deriving the corresponding MSE. Concretely, the first part is broken down into three steps. Firstly, we find the fingerprint spectrum sequence corresponding to the  asymptotically optimal window. Then, we derive the estimation MSE corresponding to the asymptotically optimal window.  
In the third step, we derive the asymptotically optimal window.  
However, since the acquisition of the asymptotically optimal window is practically unachievable, the second part of our work aims for designing  superior practical window functions using a novel insight. Bearing in mind that the CMCC algorithm can be generalized to other bi-static integrated sensing and communication (ISAC) systems, the CMCC and this window function research may also be utilized in other kind of bi-static systems, such as that of \cite{zhang2021enabling}. 
Specifically, our key contributions are summarized as follows.
\begin{itemize}
	\item[$\bullet$]
	We propose an asymptotically optimal fingerprint spectrum sequence, which achieves extremely low MSE in high-SNR scenarios. Moreover, based on the expression of the asymptotically optimal fingerprint spectrum, we derive the corresponding MSE. Both the fingerprint spectrum derived and MSE are formulated in Proposition \ref{pro1}. 
\end{itemize}
\begin{itemize}
	\item[$\bullet$]	
	We derive the mathematical relationship between the fingerprint spectrum and the  window utilized. Specifically, to achieve this in Proposition \ref{pro2}, we first derive the equation for the fingerprint spectrum sequence and the cross-correlation sequence of the  window utilized. 
	Secondly, we develop Proposition \ref{pro3} to establish the relationship between the  window utilized and the cross-correlation sequence. According to our derivation, achieving a specifically designed fingerprint spectrum relies on many parameters, which we cannot acquire before sensing. 
\end{itemize}
\begin{itemize}
	\item[$\bullet$]
	We disclose that the estimation performance for fingerprint-spectrum-based synchronization improves as the mainlobe of the Fourier transform of the window used becomes sharper. Building on this finding, we test a method that employs a well-known super-resolution estimation algorithm, specifically the MUSIC algorithm, to generate a fingerprint spectrum. Due to the high resolution and inherently sharp lobe characteristics of this algorithm, the resulting spectrum's lobe is significantly narrower than that of traditional windows, such as the rectangular and Hamming windows. Numerical simulations confirm the superior synchronization performance of the bell-shaped window with the MUSIC algorithm (BS-W-MUSIC). 
\end{itemize}

The remainder of this paper is organized as follows. In Section II, we describe the asynchronous MIMO-OFDM PMN system model and formulate the window function design problem. Then, we design practical window functions in Section III.  In Section IV, numerical simulations are conducted to verify our analysis and to confirm the superior performance of the proposed window. Finally, our conclusions are offered in Section VI.

\textit{Notations}:  We use lower-case and upper-case boldface letters to represent vectors and matrices, respectively. 
${\bf A}^{\rm T}, {\bf A}^{*}, {\bf A}^{\rm H}$ and ${\bf A}^{-1}$, ${\bf A}^{\dagger}$ represent transpose, conjugate, conjugate transpose, inverse and the pseudo-inverse of ${\bf A}$, respectively; Moreover, ${\mathrm{diag}}({\bf a}_1,\cdots,{\bf a}_n)$ is a block diagonal matrix whose diagonal blocks are $\{{\bf a}_1,\cdots,{\bf a}_n\}$. ${\bf A}[i,:]$ and ${\bf A}[:,j]$ represents the $i$th row and $j$th column of matrix ${\bf A}$; ${\bf a}[i]$ and ${\bf A}[i,j]$ represents the $i$th element of ${\bf a}$ and $(i,j)$th element of ${\bf A}$, respectively; ${a}(t)$, ${\bf a}(t)$, and ${\bf A}(t)$ are the scalar function, the vector function, and the matrix function with respect to $t$, respectively; ${\bf I}_N$ and ${\bf 0}_{M\times N}$ are the $N\times N$ identity matrix and $M\times N$ zero matrix, respectively; $\rm{Round}(\alpha)$ and ${\rm E}(\cdot)$ represent the nearest integer to $\alpha$ and the expectation operator, respectively; $\lfloor\cdot\rfloor$, ${\rm Int}(\cdot)$ and ${\rm Frac}(\cdot)$ denotes the floor function, the integer part and the decimal parts of a real number; Finally, $\oplus$, $\odot$, $\circledast$ and ${\rm mod}(\cdot)$ are the right cyclic shift operator, the Hadamard product operator, the circular convolution operator, and the modulus  operator, respectively. 

\section{Syetem Model}

\begin{figure}[tbp]
	\centering
	\includegraphics[width=3.5in]{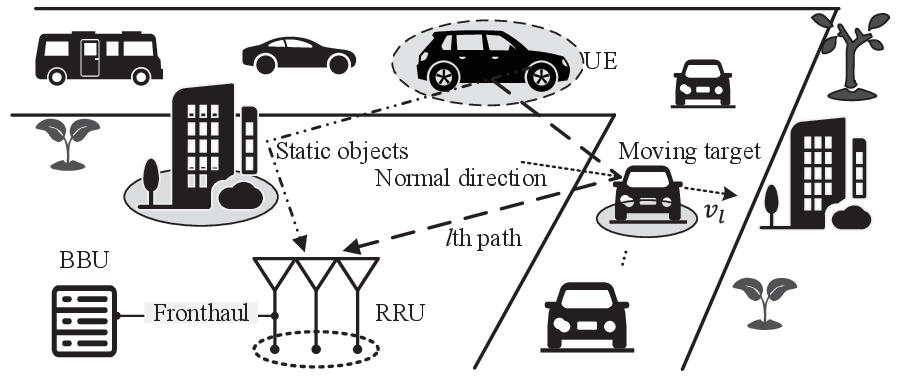}
	\caption{System model.}\label{Sketch}
\end{figure}

The PMN system is depicted in Fig. \ref{Sketch}, where a $M_{\rm r}$-antenna RRU receives signals transmitted by the UE. The propagation channel consists of $L$ paths. Let us define the propagation delay and the Doppler offset corresponding to the $l$th path as $\tau_{l}$ and $f_{{\rm D},l}$. Then, by estimating $\tau_{l}$ and $f_{{\rm D},l}$ for $l=1,\cdots,L$, the RRU achieves both velocity measurement and ranging \cite{rahman2019framework}. However, the existence of CFO, namely $\delta^f$, and TO, namely $\delta^\tau$, between the UE and the RRU will introduce ambiguity in the estimation of the desired parameters: $\tau_{l}$ and $f_{{\rm D},l}$ for $l=1,\cdots,L$ \cite{akdeniz2014millimeter}. To mitigate the ambiguity, the whole data payload can be utilized as the sensing resource to extract both delay-Doppler and CFO/TO information, as described in \cite{rahman2019framework,zhang2021enabling}.

Specifically, we denote the carrier frequency and the subcarrier spacing in the system by $f_c$ and $\Delta f$. Moreover, let us assume that there are a total of $N_{\rm c}$ OFDM subcarriers employed in the system and define $c_n$ as the data modulated on the $n$th subcarrier. 
Moreover, we denote the number of antennas of the UE as $M_{\rm t}$ and the transmit precoder (TPC) of the UE as ${\boldsymbol{\omega}}\in\mathbb{C}^{M_{\rm t}\times1}$. Therefore, we can represent the analog signal transmitted by the UE as
\begin{equation}\label{x(t)}
{\bf x}(t)=\left[e^{j2\pi f_ct}\sum_{n=0}^{N_{\rm c}-1}c_ne^{j2\pi n\Delta ft}\right]{\boldsymbol{\omega}}.
\end{equation}

Furthermore, the channel impulse response (CIR) of $l$th path can be formulated as\footnote{It's worth noting that in this work, the time-varying CIR of the $l$th path should be represented as ${\bf H}_{l}(t,\tau_l)$. However, given that $\tau_l$ is also a function of $t$, we simplify the representation to ${\bf H}_{l}(t)$.}
\begin{equation}\label{h_l^n(t)}
{\bf H}_{l}(t)=a_{l}\delta[(1-\frac{2v_l}{c})t-(\tau_{l}+\delta^\tau)]{\bf \Omega}_{l}^{\rm r}{\bf \Omega}_{l}^{\rm t},
\end{equation} where $a_l$ and $v_l$ are defined as the channel gain of the $l$th path and the velocity of the reflector projected onto the normal direction of the $l$th path, respectively. Moreover, $\delta(t)$, ${\bf \Omega}_{l}^{\rm r} =[1,e^{j2\pi d\cos(\theta_{l}^{\rm r})/\lambda},\cdots,e^{j2\pi(M_{\rm r}-1) d\cos(\theta_{l}^{\rm r})/\lambda}]^{\rm T}$, and ${\bf \Omega}_{l}^{\rm t} =[1,e^{j2\pi d\cos(\theta_{l}^{\rm t})/\lambda},\cdots,e^{j2\pi(M_{\rm r}-1) d\cos(\theta_{l}^{\rm t})/\lambda}]$ are the impulse function, the received steering vector, and the transmitted steering vector  of the signal propagated from the $l$th path, respectively. Furthermore, $d$, $\theta_{l}^{\rm r}$, $\theta_{l}^{\rm t}$, and $\lambda$ are defined as the antenna spacing, the angle of arrival (AOA), the angle of departure (AOD) corresponding to the $l$th path, and the signal wavelength, respectively. 

Then, since CP converts linear convolution to circulant convolution, the received analog baseband signal of the $g$th OFDM symbol after removing the cyclic prefix (CP) can be formulated as (\ref{y_g^n(t)}) 
\begin{equation}\label{y_g^n(t)}
\begin{aligned}
&{\bf y}_{g}(t) = \ e^{-j2\pi (f_c+\delta^f)t}\ \sum_{l=1}^{L}[{\bf H}_{l}(t)\ \circledast\ {\bf x}(t)]\\
&= \sum_{l=1}^{L}\sum_{n=0}^{N_{\rm c}-1}a_{l}c_n{\bf \Omega}_{l}^{\rm r}{\bf \Omega}_{l}^{\rm t}{\boldsymbol{\omega}}e^{-2\pi\delta^ft} e^{-j2\pi (f_c+n\Delta f)[\frac{2v_l}{c}t+(\delta^\tau+\tau_{l})]}\\
&\cdot e^{j2\pi n\Delta ft} +{\bf w}_{g}(t),
\end{aligned}
\end{equation} where ${\bf w}_{g}(t)$ represents the complex additive white Gaussian noise (AWGN) process having zero mean and variance of $\sigma^2$ for any $t$.

Upon acquiring ${\bf y}_{g}(t)$, the sampler will digitize it into a discrete-time signal. To represent the discrete-time signal, we define the sampling interval and the length of an entire OFDM symbol as $T_{\rm sam}$ and $T_{
\rm sym}$, respectively. 
Then, the $u$th sample of the digitized ${\bf y}_{g}(t)$ can be formulated as 
\begin{equation}\label{Y_g}
\begin{aligned}
&{\bf y}_{g,u}\!\!=\sum\nolimits_{l=1}^{L}\sum\nolimits_{n=0}^{N_{\rm c}-1} e^{-j2\pi f_c[\frac{2v_l}{c}(uT_{\rm sam}+(g-1)T_{\rm sym}+(\delta^\tau+\tau_{l})]}\\
&\cdot e^{-j2\pi n\Delta f\frac{2v_l}{c}(uT_{\rm sam}+(g-1)T_{\rm sym})}e^{-2\pi\delta^f(uT_{\rm sam}+(g-1)T_{\rm sym})}\\
&\cdot e^{-j2\pi n\Delta f(\delta^\tau+\tau_{l})}e^{-j2\pi n\Delta f uT_{\rm sam}}a_{l}c_n{\bf \Omega}_{l}^r{\bf \Omega}_{l}^t{\boldsymbol{\omega}}+{\bf w}_{g,u},
\end{aligned}
\end{equation}where ${\bf w}_{g,u}$ is defined as the complex AWGN vector with zero mean and covariance matrix $\sigma^2{\bf I}$, respectively. Noting that $[uT_{\rm sam}\!+\!(g-1)T_{\rm sym}]N_{\rm c}\Delta f\frac{2v_l}{c}$ is always small, one can  approximate  $e^{-j2\pi n\Delta f\frac{2v_l}{c}[uT_{\rm sam}\!+\!(g-1)T_{\rm sym}]}$ by 1 \cite{liu2020super}. Moreover, since $2\pi f_c\frac{2v_l}{c}T_{\rm sym}$ and $2\pi T_{\rm sym}\delta^f$ are also much smaller than 1, $e^{-j2\pi f_c[\frac{2v_l}{c}(uT_{\rm sam}\!+\!(g-1)T_{\rm sym})]}$ and $e^{-2\pi\delta^f[uT_{\rm sam}\!+\!(g-1)T_{\rm sym}]}$ can be approximated by $e^{-j2\pi f_c[\frac{2v_l}{c}(g-1)T_{\rm sym}]}$ \cite{liu2020super} and $e^{-2\pi\delta^f(g-1)T_{\rm sym}}$, respectively. 

By implementing these approximations, the $g$th OFDM, ${\bf Y}_g=[{\bf y}_{g,1},\cdots,{\bf y}_{g,N_{\rm c}}]$, can be further expressed as 
\begin{equation}\label{newY_g}
\begin{aligned}
{\bf Y}_g=&\sum_{l=1}^{L} e^{-j2\pi f_c[(\frac{2v_l}{c}+\frac{\delta^f}{f_c})(g-1)T_{\rm sym}]}e^{-j2\pi f_c(\delta^\tau+\tau_{l})}\\
&\cdot a_{l}{\bf \Omega}_{l}^r{\bf \Omega}_{l}^t{\boldsymbol{\omega}}{\boldsymbol\tau}_{l} {\bf D}{\bf F}\!+\!{\bf W}_g,
\end{aligned}
\end{equation}where ${\boldsymbol\tau}_{l}$ is defined as $[1,\cdots,e^{-j2\pi (N_{\rm c}-1)\Delta f(\delta^\tau+\tau_{l})}]$, ${\bf D}$ represents ${\mathrm{diag}}([c_1,\cdots,c_N])$, ${\bf F}$ is defined as the $N_{\rm c}$-dimensional inverse fast Fourier transform ($N_{\rm c}$-D IFFT) matrix, and ${\bf W}_g$ is defined as $[{\bf w}_{g,1}, \cdots, {\bf w}_{g,N_{\rm c}}]$, respectively.
To further estimate the desired parameters, namely the Doppler offset and time delay, a feasible method is to firstly compensate the received signal exploiting the demodulated communication data as 
\begin{equation}\label{compensation1}
\breve{\bf Y}_g={\bf Y}_g{\bf F}^{\rm H}{\bf D}^{-1},
\end{equation} which is also exploited in \cite{liu2020super}. Specifically, the process of utilizing data payload for target sensing is depicted in Fig. 2. The only connection between sensing synchronization studied in this paper and the communication module can be intuitively seen from the demodulated data flow. The data demodulated from the communication module is fed to the sensing module. Then, the data is first compensated and then used for our synchronization and target sensing. Given the unidirectional data flow and assuming there are no symbol demodulation errors in the flow, the performance of synchronization studied in this paper and the communication module will have no impact on each other.

\begin{figure}[h]
	\centering
	\includegraphics[width=3.5in]{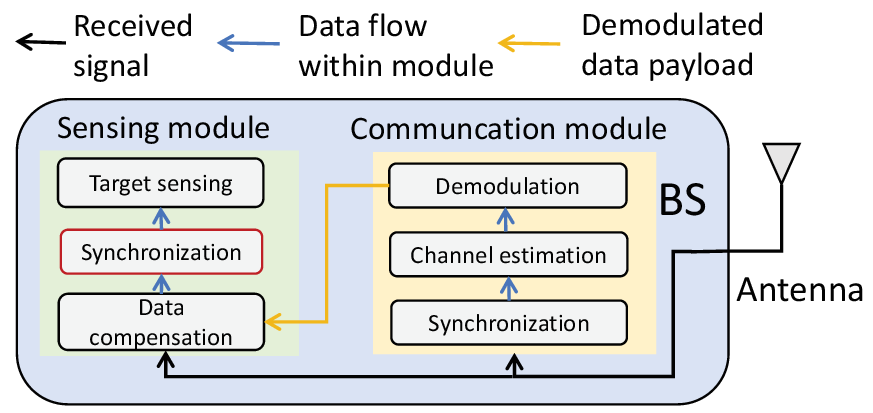}
	\caption{Data processing flow at receivers in PMN systems.}
	\label{figure3}
\end{figure}

Furthermore, let us denote the $m$th row of $\breve{\bf Y}_g$ as $\breve{\bf y}_{g,m}$. Then, we stack the received signals corresponding $G$ OFDM symbols  into $\boldsymbol{\Gamma}_m$ as $[(\breve{\bf y}_{1,m})^{\rm T},\cdots,(\breve{\bf y}_{G,m})^{\rm T}]^{\rm T}$, and assume that both the TO and the phase drift are constants within the packet. 
Specifically, $\boldsymbol{\Gamma}_m$ can be formulated as
\begin{equation}
\boldsymbol{\Gamma}_m=\sum\nolimits_{l=1}^{L}{\boldsymbol\alpha}_{l}[m]\boldsymbol{\theta}_l\boldsymbol{\tau}_l+\breve{\bf W}_m,
\end{equation} where ${\boldsymbol\alpha}_{l}=e^{-j2\pi f_c(\delta^\tau+\tau_{l})}a_{l}{\bf \Omega}_{l}^{\rm r}{\bf \Omega}_{l}^{\rm t}{\boldsymbol{\omega}}$,  $\boldsymbol{\theta}_l=[1, \cdots, e^{-j2\pi f_c(\frac{2v_l}{c}+\frac{\delta^f}{f_c})(G-1)T_{\rm sym}}]^{\rm T}$ and $\breve{\bf W}_m$ is the complex AWGN matrix, respectively.


\section{Windowing for Synchronization and Formulation of Our Optimization Problem}
In this section, we firstly extend the synchronization algorithm of \cite{wxy} by applying window functions for performance improvement. Then, we formulate an optimization problem to find the optimal window function. 

As presented in \cite{wxy}, a two-dimensional Fourier transform (2D-FFT) is firstly applied to $\boldsymbol{\Gamma}_m\in\mathbb{C}^{G\times N_c}$. We then introduce the window functions during the process of the 2D-FFT. To represent the output of the 2D-FFT, $\boldsymbol{\psi}_{G}\in {\mathbb C}^{1\times G}$ is defined as the $G$-element window function. As a result, the output ${\bf F}_G^*{\mathrm{diag}}(\boldsymbol{\psi}_{G})\boldsymbol{\Gamma}_m{\mathrm{diag}}(\boldsymbol{\psi}_{N_c}){\bf F}_{Nc}^*$ can be expressed as (\ref{2D-DFT-1}),
where $\tilde{{{\bf W}}} = {\bf F}_G^*\breve{\bf W}_m{\bf F}_{Nc}^*$ is the AWGN matrix with zero mean and covariance matrix $\tilde{\sigma}^2{\bf I}$. Then we have
\begin{equation}\label{2D-DFT-1}
\begin{aligned}
&{\bf F}_G^*{\mathrm{diag}}(\boldsymbol{\psi}_{G})\boldsymbol{\Gamma}_m{\mathrm{diag}}(\boldsymbol{\psi}_{N_c}){\bf F}_{Nc}^* \\
&=\sum\nolimits_{l=1}^{L}\!\boldsymbol{\alpha}_l[m][{\bf F}_G^{\rm *}{\mathrm{diag}}(\boldsymbol{\psi}_{G})\boldsymbol{\theta}_l][\boldsymbol{\tau}_l{\mathrm{diag}}(\boldsymbol{\psi}_{N_c}){\bf F}_{N_c}^{\rm *}]\!+\!\tilde{{\bf W}}.
\end{aligned}
\end{equation} To reduce complexity in the 2D-FFT, it is possible to substitute ${\boldsymbol{\Gamma}}_m$ with its real part \cite{wxy}. However, it is important to note that this operation leads to a half of the maximum unambiguous range.

It is important to note that the 2D-FFT can be substituted by alternative spectrum analysis methods, such as the MUSIC and ESPRIT algorithms. Hence the 2D-FFT is adopted, because it makes closed-form performance analysis tractable. A detailed performance comparison between the MUSIC and 2D-FFT methods is provided later in section \uppercase\expandafter{\romannumeral6} of this paper. Additionally, it is worth mentioning that the windows $\boldsymbol{\psi}_{G}$ and $\boldsymbol{\psi}_{{N_c}}$ used in the 2D-FFT can be of different types. For instance, $\boldsymbol{\psi}_{G}$ may be a rectangular window with $G$ elements, while $\boldsymbol{\psi}_{{N_c}}$ could be a Hamming window with $N_c$ elements. However, for the sake of brevity, we assume that $\boldsymbol{\psi}_{G}$ and $\boldsymbol{\psi}_{{N_c}}$ are of the same type in the following discussions.

To improve the estimation performance, zero-padding can be utilized \cite{oppenheim1978applications}. In the following, we implement $K$ times zero-padding. In such circumstances, (\ref{2D-DFT-1}) can be re-written as
\begin{equation}\label{newequ9}
\begin{aligned}
&{\bf F}_{KG}^*{\mathrm{diag}}(\boldsymbol{\bar\psi}_{K,G})\boldsymbol{\bar\Gamma}_{K,m}{\mathrm{diag}}(\boldsymbol{\bar\psi}_{K,N_c}){\bf F}_{KNc}^*\!=\!\tilde{{\bf W}}_K\!+\!\sum_{l=1}^{L}\!\boldsymbol{\alpha}_l[m]\\
&\![{\bf F}_{KG}^{\rm *}{\mathrm{diag}}(\boldsymbol{\bar\psi}_{K,G})\boldsymbol{\bar\theta}_{K,l}][{\bf F}_{KN_c}^{\rm H}{\mathrm{diag}}(\boldsymbol{\bar\psi}_{K,N_c}){\boldsymbol {\bar\tau}}^{\rm T}_{K,l}]^{\rm T},
\end{aligned}
\end{equation}
where we have $\boldsymbol{\bar\psi}_{K,G}=[\boldsymbol{\psi}_{G},{\bf 0}_{1\times (K-1)G}]$, $\boldsymbol{\bar\psi}_{K,N_c}=[\boldsymbol{\psi}_{N_c},{\bf 0}_{1\times (K-1)N_c}]$, $\boldsymbol{\bar\theta}_{K,l}=[\boldsymbol{\theta}^{\rm T}_{l},{\bf 0}_{1\times (K-1)G}]^{\rm T}$, ${\boldsymbol {\bar\tau}}_{K,l}=[{\boldsymbol {\tau}}_{l},{\bf 0}_{1\times (K-1)N_c}]$, and $\boldsymbol{\bar\Gamma}_{K,m}$ is defined as
\begin{equation}
\boldsymbol{\bar\Gamma}_{K,m}=\left[
\begin{matrix}
\boldsymbol{\Gamma}_m & {\bf 0}_{G\times (K-1)N_c}\\
{\bf 0}_{(K-1)G\times N_c} & {\bf 0}_{(K-1)G\times (K-1)N_c}
\end{matrix}\right].
\end{equation}
Moreover, $\tilde{{\bf W}}_K$ is defined as
\begin{equation}
\tilde{{\bf W}}_K={\bf F}_{KG}^*\left[
\begin{matrix}
\breve{\bf W}_m & {\bf 0}_{G\times (K-1)N_c}\\
{\bf 0}_{(K-1)G\times N_c} & {\bf 0}_{(K-1)G\times (K-1)N_c}
\end{matrix}\right]{\bf F}_{KN_c}^*.
\end{equation}

Then, we denote ${\bf F}_{KG}^*{\mathrm{diag}}(\boldsymbol{\bar\psi}_{K,G})\boldsymbol{\bar\Gamma}_{K,m}{\mathrm{diag}}(\boldsymbol{\bar\psi}_{K,N_c}){\bf F}_{KNc}^*$ by $\boldsymbol{\Xi}$, $[{\bf F}_{KG}^{\rm *}{\mathrm{diag}}(\boldsymbol{\bar\psi}_{K,G})\boldsymbol{\bar\theta}_{K,l}]$ by the row vector $\boldsymbol{\gamma}_{l,G}$, and $[{\bf F}_{KN_c}^{\rm H}{\mathrm{diag}}(\boldsymbol{\bar\psi}_{K,N_c}){\boldsymbol {\bar\tau}}^{\rm T}_{K,l}]^{\rm T}$ by the column vector $\boldsymbol{\gamma}^{\rm T}_{l,N_c}$. Moreover,  $\boldsymbol{\gamma}_{l,G}$ can also be expressed by the discrete-time Fourier transform (DTFT) of ${\mathrm{diag}}(\boldsymbol{\bar\psi}_{K,G})\boldsymbol{\bar\theta}_{K,l}$, namely
\begin{equation}
\boldsymbol{\bar\gamma}_{l,G}(f)=\sum\nolimits_{g=0}^{KG-1}\boldsymbol{\bar\psi}_{K,G}[g]\boldsymbol{\bar\theta}_{K,l}[g]e^{-j2\pi fg}.
\end{equation}
Moreover, to associate the continuous DTFT output $\boldsymbol{\bar\gamma}_{l,G}$ and the discrete DFT output $\boldsymbol{\gamma}_{l,G}$, the grid size of $\boldsymbol{\gamma}_{l,G}$ has to be determined. For brevity, let us define $T_{\rm R}$ and $F_{\rm R}$ as the grid size in the delay-Doppler domain, respectively. $F_{\rm R}$ can be expressed as the reciprocal of the total time period, namely as $F_{\rm R}=1/(GT_{\rm sym})$, while $T_{\rm R}$ as the reciprocal of the entire bandwidth, i.e. $T_{\rm R}=1/(N\Delta f)$, according to \cite{liu2020super}. Due to the employment of  zero-padding, the grid size actually becomes $F_{\rm R}=1/(KGT_{\rm sym})$ and $\frac{T_{\rm R}}{K}$ \cite{oppenheim1978applications}. Thus, we have
$\boldsymbol{\gamma}_{l,G}[i]=\boldsymbol{\bar\gamma}_{l,G}({iF_{\rm R}})=\boldsymbol{\bar\gamma}_{l,G}(\frac{i \Delta f}{KG})$ and $
\boldsymbol{\gamma}^{\rm T}_{l,N_c}[p]=\boldsymbol{\bar\gamma}_{l,N_c}({pT_{\rm R}})=\boldsymbol{\bar\gamma}_{l,N_c}(\frac{pT_{\rm sam}}{K}).
$
Similarly, the composition of $\boldsymbol{\gamma}^{\rm T}_{l,N_c}$ is consistent with that of $\boldsymbol{\gamma}_{l,G}$.
As a result, $\boldsymbol{\Xi}[i,p] $ can be expressed as
\begin{equation}\label{xi}
\begin{aligned}
\boldsymbol{\Xi}[i,p]\!=\!\! \sum_{l=1}^{L}2\pi^2\boldsymbol{\alpha}_l[m]\boldsymbol{\bar\gamma}_{l,N_c}\!\!\left(\!\frac{pT_{\rm sam}}{K}\!\right)\boldsymbol{\bar\gamma}_{l,G}\left(\!\frac{i \Delta f}{KG}\right)\!\!+\!\!{\tilde{\bf W}}_K[i,\!p].
\end{aligned}
\end{equation} 

Although the closed-form expression of $\boldsymbol{\Xi}$ has been obtained, it is not straightforward to see its shape. We can further reformulate the expression of $\boldsymbol{\Xi}$ as follows.
According to (\ref{newequ9}), we know the term $[{\bf F}_{KG}^{\rm *}{\mathrm{diag}}(\boldsymbol{\bar\psi}_{K,G})\boldsymbol{\bar\theta}_{K,l}]$. Furthermore, $[{\bf F}_{KN_c}^{\rm H}{\mathrm{diag}}(\boldsymbol{\bar\psi}_{K,N_c}){\boldsymbol {\bar\tau}}^{\rm T}_{K,l}]^{\rm T}$ represents the Fourier transforms of the product of ${\mathrm{diag}}(\boldsymbol{\bar\psi}_{K,G})$ and $\boldsymbol{\bar\theta}_{K,l}$, and ${\mathrm{diag}}(\boldsymbol{\bar\psi}_{K,N_c})$ and ${\boldsymbol {\bar\tau}}^{\rm T}_{K,l}$. According to the convolution theorem \cite{oppenheim1978applications}, namely the Fourier transform of a convolution between two sequences equals the product of the Fourier transforms of the two sequences, $[{\bf F}_{KG}^{\rm *}{\mathrm{diag}}(\boldsymbol{\bar\psi}_{K,G})\boldsymbol{\bar\theta}_{K,l}]$ and $[{\bf F}_{KN_c}^{\rm H}{\mathrm{diag}}(\boldsymbol{\bar\psi}_{K,N_c}){\boldsymbol {\bar\tau}}^{\rm T}_{K,l}]^{\rm T}$ can be re-formulated as
\begin{equation}\label{equ9}
\begin{aligned}
&\boldsymbol{\gamma}_{l,G}=[({\bf F}_{KG}^{\rm *}\boldsymbol{\bar\psi}_{K,G}^{\rm T})\!\boxtimes\!({\bf F}_{KG}^{\rm *}\boldsymbol{\bar\theta}_{K,l})],\\
&\boldsymbol{\gamma}^{\rm T}_{l,N_c}=[({\bf F}_{KN_c}^{\rm H}\boldsymbol{\bar\psi}_{K,N_c}^{\rm T})\!\boxtimes\!({\bf F}_{KN_c}^{\rm H}{\boldsymbol {\bar\tau}}^{\rm T}_{K,l})]^{\rm T},
\end{aligned}
\end{equation}
where $(\boxtimes)$ represents a matrix-wise convolution operator, defined as $[{\bf a}_1, \cdots, {\bf a}_i]\boxtimes[{\bf b}_1, \cdots, {\bf b}_i]=[{\bf a}_1\circledast{\bf b}_1,
\cdots, {\bf a}_i\circledast{\bf b}_i]$. Here, ${\bf a}_i$ and ${\bf b}_i$ represent any column vectors of the same dimension. It is plansible  that $\boldsymbol{\gamma}_{l,G}$ represents the discrete circular convolution sequence between the Fourier transform of the window function $\boldsymbol{\bar\psi}_{K,G}$ and  the Fourier transforms of $\boldsymbol{\bar\theta}_{K,l} $.  Moreover, since $\boldsymbol{\bar\theta}_{K,l} $ is the single-tone discrete sequence, $\boldsymbol{\bar\gamma}_{l,G}(f)$ can be naturally expressed as the cyclic-shift based Fourier transform of $\boldsymbol{\bar\psi}_{K,G}$. The symmetry axis of $\boldsymbol{\bar\gamma}_{l,G}(f)$ is at the exact frequency of the sequence $\boldsymbol{\bar\theta}_{K,l} $, namely at $f=\delta^f+f_{{\rm D},l}$. Similarly, the symmetry axis of $\boldsymbol{\bar\gamma}_{l,N_c}(f)$ is located at $f=\delta^\tau+\tau_l$.

Furthermore, by substituting (\ref{equ9}) into (\ref{newequ9}) we obtain
\begin{equation}\label{2D-DFT-3}
\begin{aligned}
{\boldsymbol{\Xi}}=&\tilde{{\bf W}}_K+\sum\nolimits_{l=1}^{L}\boldsymbol{\alpha}_l[m][({\bf F}_{KG}^{\rm *}\boldsymbol{\bar\psi}_{K,G}^{\rm T})\boxtimes({\bf F}_{KG}^{\rm *}\boldsymbol{\bar\theta}_{K,l})]\\
&[({\bf F}_{KN_c}^{\rm H}\boldsymbol{\bar\psi}_{K,N_c}^{\rm T})\boxtimes({\bf F}_{KN_c}^{\rm H}{\boldsymbol {\bar\tau}}^{\rm T}_{K,l})]^{\rm T}.
\end{aligned}
\end{equation}
In the matrix $\boldsymbol{\Xi}$, the elements in each row and column are acquired by sampling from the same function $\bar{\boldsymbol\gamma}_{l,G}(f)$ and $\bar{\boldsymbol\gamma}_{l,N_c}(f)$, respectively. As a result, it is intuitive that $\boldsymbol{\Xi}$ is the sum of the Gaussian noise and the linear summation of $L$ discrete 2-D sample matrices, namely $\boldsymbol{\gamma}_{l,G}\boldsymbol{\gamma}^{\rm T}_{l,N_c}$ for $l=1,\cdots,L$ \cite{singh2008communication}, whose centers are at $[(\delta^f+f_{{\rm D},l})/F_{\rm R} ,(\delta^\tau+\tau_l)/{T_{\rm R}}]$. Each of these discrete 2D sample matrices, $\boldsymbol{\gamma}_{l,G}\boldsymbol{\gamma}^{\rm T}_{l,N_c}$, corresponds to signals  reflected by objects. Among the $L$ objects, we assume that there exist $L_1$ static objects in the environment. Since they are static, $ f_{{\rm D},l}$ and $\tau_{{l}}$ corresponding to these $L_1$ objects are 0 and constant, respectively. Thus, according to (\ref{xi}), when the normalized CFO, $\delta^f$, and TO, $\delta^\tau$, drift, these discrete 2D sample matrices corresponding to the $L_1$ static objects will consequently change. Let us consider the $(i,p)$th element of $\boldsymbol{\gamma}_{l,G}\boldsymbol{\gamma}^{\rm T}_{l,N_c}$ as an example, which is given by ${\boldsymbol{\bar\gamma}}_{l,G}\left[\frac{i \Delta f}{KG}\right]{\boldsymbol{\bar\gamma}}_{l,N_c}\left[\frac{pT_{\rm sam}}{K}\right]$. If the CFO and TO are increased by $\Delta\delta^f$ and $\Delta\delta^\tau$ respectively, the transformed expression becomes ${\boldsymbol{\bar\gamma}}_{l,G}\left[\frac{i \Delta f}{KG}+\Delta\delta^f\right]{\boldsymbol{\bar\gamma}}_{l,N_c}\left[\frac{pT_{\rm sam}}{K}+\Delta\delta^\tau\right]$.
In this context, it is easy to see that the sampling points, corresponding to each sample of these sample matrices, in the 2D function ${\boldsymbol{\bar\gamma}}_{l,G}(f){\boldsymbol{\bar\gamma}}_{l,N_c}(f)$ will undergo circular translation.

Moreover, it is natural that the symmetry points of these $L_1$ 2D functions are $(\delta^f/F_{\rm R},(\delta^\tau+\tau_l)/{T_{\rm R}})$ because $f_{{\rm D},l}=0$ for these static objects, which means that these 2D functions are located at the same row index. Consequently, we have: $\boldsymbol{\bar\gamma}_{l,G}(f)=\boldsymbol{\bar\gamma}_{1,G}(f)$ and $\boldsymbol{\gamma}_{l,G}=\boldsymbol{\gamma}_{1,G}$ for $l=1,\cdots,L$. As a result, by locating the row index of the symmetry points of these $L_1$ 2D functions, which is assumed as $K_0$, we can define ${\boldsymbol{\beta}}_{\rm n}=\boldsymbol{\Xi}[K_0,:]$ as the \textit{fingerprint spectrum}. The fingerprint spectrum sequence contains the location distribution information of these $L_1$ static objects in the environmental background.  
Essentially, the sequence may be deemed to be an identity code of the specific static environment. 
However, ${\boldsymbol{\beta}}_{\rm n}$ is the fingerprint spectrum polluted by noise. For simplicity, we here define ${\boldsymbol\beta}\in\mathbb{C}^{1\times KN_c}$ as the pure fingerprint spectrum vector without noise and ${\boldsymbol{\beta}}$ can be formulated as
\begin{equation}
{\boldsymbol{\beta}}={\boldsymbol{\beta}}_{\rm n}-\tilde{\bf W}_K[K_0,:].
\end{equation}

Then, if the CFO increases by $\Delta\delta^f$ and the TO increases by $\Delta\delta^\tau$, the offset fingerprint spectrum contaminated by noise can be defined as ${\boldsymbol{\beta}}_{\rm n, c}=\boldsymbol{\Xi}[K_0+{\rm Round}(\Delta\delta^f/F_{\rm R}),:]$. Moreover, we also define ${\boldsymbol{\beta}}_{\rm c}$ as the offset fingerprint spectrum without noise. Therefore, by assuming that $\Delta\delta^f/F_{\rm R}$ and $\frac{\Delta\delta^\tau}{T_{\rm R}}$ are integers, the relationship between ${\boldsymbol{\beta}}$ and ${\boldsymbol{\beta}}_{\rm c}$ can be simply represented as\footnote{Assume the location of the UE does not change or changes very slowly.} 
\begin{equation}\label{circular}
\begin{small}
\begin{aligned}
&{\boldsymbol{\beta}}_{\rm c}[q]={\boldsymbol{\beta}}\left[q+\frac{\Delta\delta^\tau}{T_{\rm R}}\right].
\end{aligned}
\end{small}
\end{equation}If $\Delta\delta^f/F_{\rm R}$ and $\frac{\Delta\delta^\tau}{T_{\rm R}}$ are not integers, we have
\begin{small}
\begin{equation}
\begin{aligned}
&{\boldsymbol{\beta}}_{\rm c}[q]\!\!=\!\!\sum_{l=1}^{L}\boldsymbol{\bar\gamma}_{l,G}(\frac{K_0 \Delta f}{KG}\!\!+\!\!\left[{\rm Int}(\frac{\Delta\delta^f\Delta f}{KG})\!\!+\!\!{\rm Fra}(\frac{\Delta\delta^f\Delta f}{KG})\right]\!\!\frac{KG}{\Delta f})\\
&\cdot\boldsymbol{\bar\gamma}_{l,N_c}(\frac{qT_{\rm sam}}{K}+\left[{\rm Int}(\frac{\Delta\delta^\tau T_{\rm sam}}{K})+{\rm Fra}(\frac{\Delta\delta^\tau T_{\rm sam}}{K})\right]\frac{K}{T_{\rm sam}})
\end{aligned}.
\end{equation}\end{small}Moreover, since ${\rm Fra}(\frac{\delta^f\Delta f}{KG})\frac{KG}{\Delta f}$ and ${\rm Fra}(\frac{\Delta\delta^\tau T_{\rm sam}}{K})\frac{K}{T_{\rm sam}}$ are always less than $\frac{\Delta f}{KG}$ and $\frac{T_{\rm sam}}{K}$, they are usually sufficiently small so that  ${\boldsymbol{\beta}}_{\rm c}[q]$ can be approximated as
\begin{equation}\label{circular2}
\begin{small}
\begin{aligned}
&{\boldsymbol{\beta}}_{\rm c}[q]\approx{\boldsymbol{\beta}}\left[q+\frac{\Delta\delta^\tau}{T_{\rm sam}}\right]=\sum_{l=1}^{L}\boldsymbol{\bar\gamma}_{l,G}(\frac{K_0 \Delta f}{KG}\!+\!\left[{\rm Int}(\frac{\Delta\delta^f\Delta f}{KG})\right]\\
&\frac{KG}{\Delta f})\cdot\boldsymbol{\bar\gamma}_{l,N_c}(\frac{qT_{\rm sam}}{K}+\left[{\rm Int}(\frac{\Delta\delta^\tau T_{\rm sam}}{K})\right]\frac{K}{T_{\rm sam}}).
\end{aligned}
\end{small}
\end{equation}

\begin{figure}[tbp]
	\centering
	\includegraphics[width=3.5in]{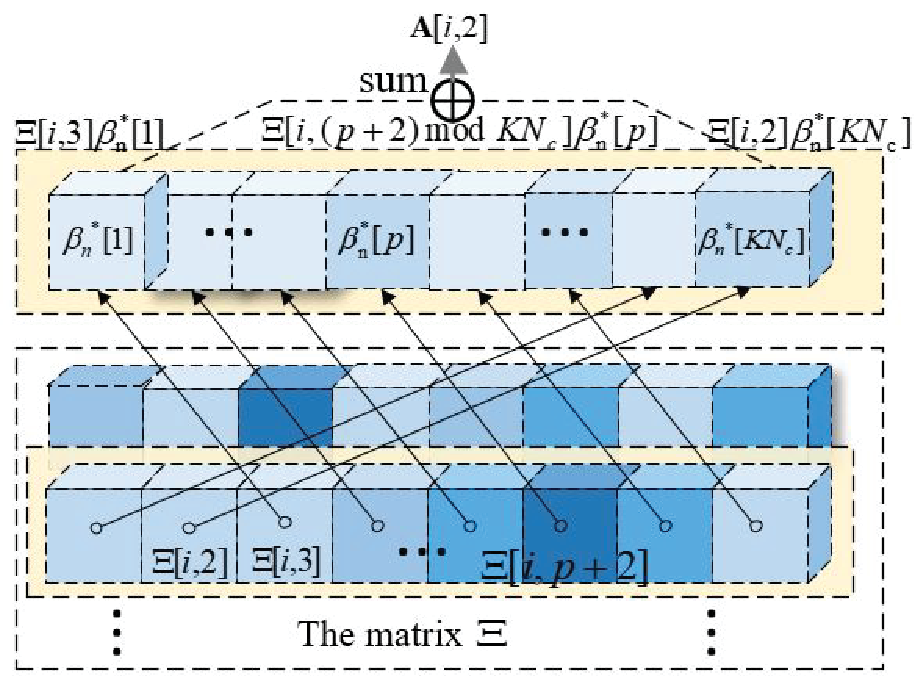}
	\caption{Demonstration of the sliding window cross-correlation in (\ref{unsimplified}).}\label{metric}
\end{figure}

Equations (\ref{circular}) and (\ref{circular2}) suggest that ${\boldsymbol{\beta}}_{\rm c}$ may be deemed to be the cyclic-shift ${\boldsymbol{\beta}}$, regardless whether $\Delta\delta^f/F_{\rm R}$ and $\frac{\Delta\delta^\tau}{T_{\rm R}}$ are integers or not. As a result, the sliding window cross-correlation analysis between ${\boldsymbol{\beta}}_{\rm n}\in\mathbb{C}^{1\times KN_c}$ and $\boldsymbol{\Xi}\in\mathbb{C}^{KG\times KN_c}$ can be implemented as (\ref{unsimplified}) in order to find the number of cyclic shifts and consequently estimate the CFO and TO. This is represented as
\begin{equation}\label{unsimplified}
\begin{aligned}
\{&\frac{\Delta\hat\delta^f}{F_{\rm R}},\frac{\Delta\hat\delta^\tau}{T_{\rm R}}\}=\mathop{\rm max}\limits_{i,q}\big|\sum_{p=1}^{{KN_c}}\frac{{\boldsymbol{\Xi}}[i,\bar q]{\boldsymbol{\beta}_{\rm n}}^*[p]}{|{\boldsymbol{\beta}_{\rm n}}^*|^2}\big|,\\
&{\rm for}\ i=1,\cdots,KG\ {\rm and} \ q=1,\cdots,KN_c,
\end{aligned}
\end{equation}
where we have $\bar q=[(q+p)\ {\rm mod}\ {KN_c}]$. Based on Equation (\ref{unsimplified}), it is evident that $\frac{\Delta\hat\delta^f}{F_{\rm R}}$ and $\frac{\Delta\hat\delta^\tau}{T_{\rm R}}$ represent the relative CFO and TO with respect to a specific time instant. To obtain the absolute values of CFO and TO, it is necessary to establish an anchor with prior knowledge of velocity and range, as assumed in \cite{IndoTrack, ni2021uplink}. For instance, this anchor could be a stationary UE in LOS scenarios or a static reflector in NLOS scenarios. However, in scenarios lacking an available anchor, the fingerprint spectrum can still be located through the following methods.
1) In scenarios with a priori UE, as outlined in \cite{ni2021uplink}, one can initially acquire the row index in LOS scenarios and subsequently utilize this index immediately upon transitioning from LOS to NLOS scenarios. Given the relatively minor change in CFO during such a brief period (the duration it takes for the scenario to shift from LOS to NLOS), the fingerprint spectrum can be accurately located.
2) In scenarios without any priori UE, one can identify the fingerprint spectrum by determining the row in $\boldsymbol{\Xi}$ with the highest power, as signals reflected by stationary objects typically exhibit greater prominence in the received signals. 
Moreover, in scenarios where part of static reflectors are prone to movement, it becomes crucial to frequently update the fingerprint spectrum and conduct (20). The frequent updating ensures that any changes resulting from the displacement of moving objects are captured continuously. Similarly, regular updates to (20) are necessary to estimate the CFO and TO under a slightly changing fingerprint spectrum, thereby maintaining estimation accuracy.

For simplicity, let us define $\sum_{p=1}^{{KN_c}}{{\boldsymbol{\Xi}}[i,\bar p]{\boldsymbol{\beta}_{\rm n}}^*[q]}/{|{\boldsymbol{\beta}_{\rm n}}^*|^2}$ as ${\bf A}[i,q]$, and by substituting (\ref{xi}) and ${\boldsymbol{\beta}_{\rm n}}$ into (\ref{unsimplified}), ${\bf A}[i,q]$ can be further represented as
\begin{small}
\begin{equation}\label{correlation}
\begin{aligned}
&{\bf A}[i,q]\!\!=\! 4\pi^4\!\frac{{\boldsymbol\gamma}_{1,G}\!\left[i\right]\!{\boldsymbol\gamma}_{1,G}\!\left[\! K_0\!\right]}{{|{\boldsymbol{\beta}_{\rm n}}^*|^2}}\!\sum_{l^{'}=1}^{L}\!\sum_{l=1}^{L}\boldsymbol{\alpha}_l[m]\boldsymbol{\alpha}_{l^{'}}[m]{\boldsymbol\rho}_{l,l^{'}}[q]\!\!+\!\!{\bf \bar W},
\end{aligned}
\end{equation}\end{small}where ${\bf \bar W}$ is the complex AWGN matrix with zero mean and variance $\bar\sigma^2$. Furthermore, ${\boldsymbol\rho}_{l,l^{'}}\in \mathbb{C}^{1\times KN_c}$ is the circular convolution of the Fourier transform of the window function, ${\boldsymbol{\gamma}}_{l,N_c}$. Moreover, ${\boldsymbol\rho}_{l,l^{'}}[q]$ can be formulated as
\begin{equation}\label{rho}
\begin{small}
{\boldsymbol\rho}_{l,l^{'}}[q]\!=\!\sum\nolimits_{p=1}^{{KN_c}}\!\!{{\boldsymbol{\bar\gamma}}_{l,N_c}\!\left[[(p+q)\ {\rm mod}\ KN_c]T_{\rm R}\right]\!{\boldsymbol{\bar\gamma}}_{l^{'},N_c}\left[pT_{\rm R}\right]}.
\end{small}
\end{equation}

From (\ref{correlation}) and (\ref{rho}), we find that ${\bf A}[i, q]$ is strongly coupled with ${\boldsymbol\gamma}^{\rm T}_{l,N_c}$ for $l=1,\cdots,L$, while ${\boldsymbol\gamma}_{l,G}$ only affects the signal-to-noise ratio (SNR) of ${\bf A}[i, q]$. 
Thus, the impact of $k$ on the estimation performance is equivalent to that of the SNR. As a result, we here set the row index $k$ as a constant $K_1={\rm Round}(\Delta\delta^f/F_{\rm R})$, which makes ${\boldsymbol{\gamma}}_G\left[ K_1\right]{\boldsymbol{\gamma}}_G\left[K_0\right]$ as large as possible for simplicity. 
 
As a result, the MSE of the delay estimate in (\ref{unsimplified}) is a function of $\boldsymbol{\psi}_{N_c}$:   
\begin{equation}\label{theoretical_MSE}
\begin{small}
\begin{aligned}
f_{\rm MSE}(\boldsymbol{\psi}_{N_c}) = \sum\nolimits_{q=1}^{KN_c}{\rm P}_{\{q,KN_c\}}(q-\frac{\Delta\delta^\tau}{T_{\rm R}})^2,
\end{aligned}
\end{small}
\end{equation} where ${\rm P}_{\{q,KN_c\}}$ is the probability that the $q$th element of the set $\{{\bf A}[K_1,q]| q=1,\cdots,K{N_c}\}$ is the maximum. According to the definition, ${\rm P}_{\{q,KN_c\}}$ can be computed as the probability distribution function (CDF) of a $KN_c$-D Gaussian random variable and it can be formulated as
\begin{equation}\label{newequ15}
\begin{small}
\begin{aligned}
&{\rm P}_{\{q,KN_c\}} = \int_{\mathbb V}p({\bf s},\tilde\sigma^2){\mathrm d}\ {\bf b}\\
&=\int_{-\infty}^{\infty} \underbrace{\int_{-\infty}^{b_q}\cdots\int_{-\infty}^{b_q}}_{KN_c-1} \frac{e^{-{ ({\bf b}-{\bf s})^T({\bf b}-{\bf s})}/{2\tilde\sigma^2}}}{(2\pi)^{\frac{d}{2}}\tilde\sigma^{d}}{\mathrm d}\ b_1\cdots {\mathrm d}\ b_{KN_c}{\mathrm d}\ b_q,
\end{aligned}
\end{small}
\end{equation}where $p({\bf s},\tilde\sigma^2)$ is the probability of the multidimensional Gaussian function with mean ${\bf s}$ and covariance matrix $\tilde\sigma^2{\bf I}_{KN_c}$. Furthermore,  ${\bf s}$ is the expectation of the $K_1$th row of ${\bf A}$. Note that A is the linear weighted sum of the circular convolution of ${\boldsymbol{\gamma}}_{N_c}$ according to (\ref{correlation}) and (\ref{rho}), while ${\boldsymbol{\gamma}}_{N_c}$ is the Fourier transform of the selected window function. As a result, the window function $\boldsymbol{\psi}_{N_c}$ has a strong influence on the synchronization MSE. 
Moreover, ${\bf b}=(b_1,\cdots, b_{KN_c})$ in (\ref{newequ15}) belongs to the $KN_c$-element vector space $\mathbb V$, which satisfies:
\begin{equation}
b_q>b_i, {\rm for}\ i\in\{1,\cdots,KN_c\}\ {\rm and}\ i\neq q.
\end{equation}



In the following, we intend to optimize the window function to minimize $f_{\rm MSE}(\boldsymbol{\psi}_{N_c})$:
\begin{equation}
\label{P}
\begin{aligned} 
\mathop{\rm min}\limits_{\boldsymbol{\psi}_{N_c}}\ & f_{\rm MSE}(\boldsymbol{\psi}_{N_c}), \\
s.t.\quad 
&\boldsymbol{\psi}_{N_c}\in\mathbb{C}^{1\times KN_c}.
\end{aligned}
\end{equation}

\section{Asymptotically Optimized Ideal Window Function}
We intend to find the solution of (\ref{P}) by the following logic.
Firstly, in Proposition 1, we derive an asymptotic optimal solution of the following problem:
\begin{small}
\begin{equation}
\label{P1}
\begin{aligned} 
\mathop{\rm min}\limits_{{{g }(\boldsymbol{\psi})}}\ & f_{\rm MSE}({{g }(\boldsymbol{\psi})}), \\
s.t.\quad 
&{{g }(\boldsymbol{\psi})}\in\mathbb{C}^{1\times KN_c}\\
&{\boldsymbol{\psi}\in\mathbb{C}^{1\times KN_c},}
\end{aligned}
\end{equation}\end{small}where ${g }(\boldsymbol{\psi})$ is defined as the function of the expectation of the $K_1$th row of ${\bf A}$, namely ${\bf s}$, with respect to the window function, $\boldsymbol{\psi}$. Moreover, Proposition \ref{pro1} also derives the synchronization MSE corresponding to the asymptotic optimal solution.
Then, based on the relation between ${\bf s}={{g }(\boldsymbol{\psi})}$ and the window function $\boldsymbol{\psi}$, Proposition \ref{pro2} 
formulates the cross-correlation of the Fourier transform of the window function utilized for any fingerprint spectrum. Finally, based on the cross-correlation derived, Proposition \ref{pro3} specifies the approximately optimal  $\boldsymbol{\psi}_{N_c}$ corresponding to any ${\bf s}$. 

Concretely, Proposition 1 is described as:

\begin{proposition}
	 If $\frac{\Delta\delta^\tau}{T_{\rm R}}$ is an integer, $f_{\rm MSE}(\boldsymbol{\psi}_{N_c})$ is asymptotically  minimized when the SNR tends to infinity and ${\bf s}$ satisfies
	 \begin{equation}\label{ideal}
	 \begin{cases}
	 {\bf s}[q]=1, q=\frac{\Delta\delta^\tau}{T_{\rm R}},\\
	 {\bf s}[q]=0, q\neq\frac{\Delta\delta^\tau}{T_{\rm R}},
	 \end{cases}
	 \end{equation} 
	 and the corresponding MSE is $f_{\rm MSE}(\boldsymbol{\psi}_{N_c})=\sum_{\substack{q=1, q\neq {\Delta\delta^\tau}/{T_{\rm R}}}}^{KN_c} $ $\left[ q_{\rm r}- \frac{\Delta\delta^\tau}{T_{\rm R}}\right]^2\int_{-\infty}^{\infty} \big[f_{\mathcal N}(b_q)$ $Q(\frac{b_q-1}{\bar\sigma^2})$ $Q^{(KN_c-2)} $ $(\!{b_q}\!/\!{\bar\sigma^2}\!){\mathrm d}\big] b_q$, where $Q(\cdot)$ is the cumulative distribution function of the standard Gaussian random variable.
	 \label{pro1}
\end{proposition}

\textit{Proof}: See Appendix \ref{proof1}.

However, unlike the hypothesis in Proposition \ref{pro1}, $\frac{\Delta\delta^\tau}{T_{\rm R}}$ can hardly be an integer in actual PMN systems. According to the criterion described in the proof of Proposition 1, to make  $f_{\rm MSE}(\boldsymbol{\psi}_{N_c})$ as small as possible, one should assign as much weight as possible to the index closest to $\frac{\Delta\delta^\tau}{T_{\rm R}}$ and as low weight as possible to indices far from $\frac{\Delta\delta^\tau}{T_{\rm R}}$. As a result, the probability corresponding to ${\rm Round}(\frac{\Delta\delta^\tau}{T_{\rm R}})$ should be assigned as large weight as possible. Under this case, ${\bf s}_{\rm ap}$ is set as
\begin{equation}\label{s_q}
\begin{cases}
{\bf s}_{\rm ap}[q]=1, q={\rm Round}(\frac{\Delta\delta^\tau}{T_{\rm R}}),\\
{\bf s}_{\rm ap}[q]=0, q\neq{\rm Round}(\frac{\Delta\delta^\tau}{T_{\rm R}}),
\end{cases}
\end{equation}
Similarly to the case where $\frac{\Delta\delta^\tau}{T_{\rm R}}$ is an integer, ${\bf s}_{\rm ap}$ in (\ref{s_q}) will be near-optimal, when $1/\tilde{\sigma}^2$ becomes large.
Specifically, based on (\ref{P_CDF1}), under this case we further derive ${\rm P}_{\{q,KN_c\}}$ by dividing $q=1,\cdots,KN_c$ into two categories, namely $q={\rm Round}(\frac{\Delta\delta^\tau}{T_{\rm R}})$ and $q\neq{\rm Round}(\frac{\Delta\delta^\tau}{T_{\rm R}})$
\begin{equation}\label{equ28}
\begin{cases}
\int_{-\infty}^{\infty}f_{\mathcal N}({b_q}/{\bar\sigma^2})Q^{(KN_c-1)}({b_q}/{\bar\sigma^2}){\mathrm d}\ b_q,\ q={\rm Round}(\frac{\Delta\delta^\tau}{T_{\rm R}}),\\
\int_{-\infty}^{\infty}\!\!f_{\mathcal N}(b_q)Q(\frac{b_q-1}{\bar\sigma^2})Q^{(KN_c-2)}(\!{b_q}\!/\!{\bar\sigma^2}\!){\mathrm d} b_q,q\!\neq\!{\rm Round}(\frac{\Delta\delta^\tau}{T_{\rm R}}).
\end{cases}
\end{equation}
Furthermore, since expressions for $q$ belong to two categories, we substitute (\ref{equ28}) into (\ref{theoretical_MSE}). Then, $f_{\rm MSE}$ can be further expressed as (\ref{newequ29}), where $q_{\rm r}$ is defined as ${\rm Round}( \frac{\Delta\delta^\tau}{T_{\rm R}})$.

\begin{figure*}[h]
	\begin{equation}
		\begin{small}
		\begin{aligned}
		&f_{\rm MSE}(\boldsymbol{\psi}_{N_c})= \left[q_{\rm r}-\frac{\Delta\delta^\tau}{T_{\rm R}}\right]^2\Big[\int_{-\infty}^{\infty}f_{\mathcal N}(b_q)Q^{(KN_c-1)}(\frac{b_{q_{\rm r}}}{\bar\sigma^2}){\mathrm d}\ b_{q_{\rm r}}
		+\sum_{\substack{q=1,\\ q\neq q_{\rm r}}}^{KN_c}\int_{-\infty}^{\infty}f_{\mathcal N}(b_q)Q(\frac{b_q-1}{\bar\sigma^2})Q^{(KN_c-2)}(\frac{b_q}{\bar\sigma^2}){\mathrm d}\ b_{q}\Big].
		\end{aligned}\label{newequ29}
		\end{small}
		\end{equation}
\end{figure*}
%

Although Proposition 1 describes the near-optimal ${\bf s}$ and the corresponding $f_{\rm MSE}(\boldsymbol{\psi}_{N_c})$, we have not obtained the expression of ${\boldsymbol\rho}_{l,l^{'}}$ corresponding to the near-optimal ${\bf s}_{\rm ap}$. Essentially, ${\bf s}_{\rm ap}$ is constituted by the circular auto-correlation sequence of the discrete Fourier transform of the selected window functions, namely ${\boldsymbol\rho}_{l,l^{'}}$ for $l=1,\cdots,L$ and $l^{'}=1,\cdots,L$. Moreover, 
for any $l$ and $l^{'}$, ${\boldsymbol\rho}_{l,l^{'}}$ is a circular-shifted version of ${\boldsymbol\rho}_{l,l}$ and for any $l$ we have: ${\boldsymbol\rho}_{1,1}={\boldsymbol\rho}_{l,l}$ according to (\ref{rho}). Thus, we intend to derive the exact ${\boldsymbol\rho}_{1,1}$ in the following. 

Then, Proposition \ref{pro2} outlines the closed-form expression of the desired ${\boldsymbol\rho}_{1,1}$. 

\begin{proposition}\label{pro2}
The $\boldsymbol{\rho}_{1,1}$ corresponding to any achievable ${\bf s}$ is formulated as
\begin{equation}
\boldsymbol{\rho}_{1,1}={{\bf s}\breve{\bf J}_{\rm Inv}},
\end{equation}  where $\breve{\bf J}_{\rm Inv}$ is defined in (\ref{newequ51})
of Appendix \ref{proof2}.
\end{proposition}

\textit{Proof}: See Appendix \ref{proof2}.




However, although Proposition \ref{pro2} derives $\boldsymbol{\rho}_{1,1}$, we still do not know the expression of $\boldsymbol{\psi}_{N_c}$. To find the window function corresponding to $\boldsymbol{\rho}_{1,1}$, we intend to further derive $\boldsymbol{\psi}_{N_c}$ based on the closed-form expression of $\boldsymbol{\rho}_{1,1}$, given by Proposition \ref{pro2}. 

\begin{proposition}\label{pro3}
	The $\boldsymbol{\bar{\psi}}_{N_c}$ corresponding to any achievable ${\bf s}$ must satisfy the condition 
	\begin{equation}\label{newequ32}
	\begin{cases}
	{\boldsymbol{\bar{\psi}}}_{K,N_c}[i]\sum_{j=1}^{N_c}{\boldsymbol{\bar{\psi}}}_{K,N_c}[j]\bar{\bf E}[i,j]\!=\!\boldsymbol{\rho}_{1,1}[i], {\rm for}\ i=1,\cdots,N_c,\\
	\boldsymbol{\rho}_{1,1}[i]=0,\ {\rm for}\ i=N_c+1,\cdots,KN_c,
	\end{cases}
	\end{equation}where $\bar{\bf E}={\bf F}_{KN_c}{\bf E}{\bf F}_{KN_c}^{\rm H}$ and ${\bf E}$ is defined as the anti-diagonal matrix, where the elements on the anti-diagonal are all 1.
\end{proposition}

\textit{Proof}: See Appendix \ref{proof3}.

According to (\ref{equ32}), the value of the globally optimal ${\boldsymbol \psi_{N_c}}$ is strongly related to both ${\boldsymbol \phi}_m$ and to time delay of every paths. Moreover, ${\boldsymbol \phi}_m$ is composed of the complex gain of every propagation path, which is hard to acquire before successful sensing. Thus, even though there exists a globally optimal $\boldsymbol{\psi}_{N_c}$, we cannot practically find it since it is strongly time-delay-and-complex-gain-dependent, before we obtain the time delay and complex gain of all paths.  

\section{Practical Window Design based on Super Resolution Estimation Algorithm}

Even though the asymptotically optimal window cannot be practically deployed, a large number of feasible discrete windows belonging to $\mathbb{C}^{1\times KN_c}$ can be selected. However, there are no mathematical criteria clarifying how to choose 'good' windows from the set of the massive variety of feasible windows $\boldsymbol{\psi}_{N_c}$. 

To establish a criterion, we revisit Proposition \ref{pro1}. According to this proposition, the cross-correlation sequence of the fingerprint spectrum follows a complex Gaussian distribution with a mean of $\sum_{p=1}^{KN_c}{\boldsymbol{\beta}}[p]{\boldsymbol{\beta}}_{\rm c}[p+q]$ for $q=1,\cdots,KN_c$, and a variance of $\bar\sigma^2$. Additionally, when $\tilde\sigma^2$ is known, $\bar\sigma^2$ is determined by $|{\boldsymbol{\beta}_{\rm n}}|^2$.

\begin{figure}[tbp]
	\centering
	\includegraphics[width=3.5in]{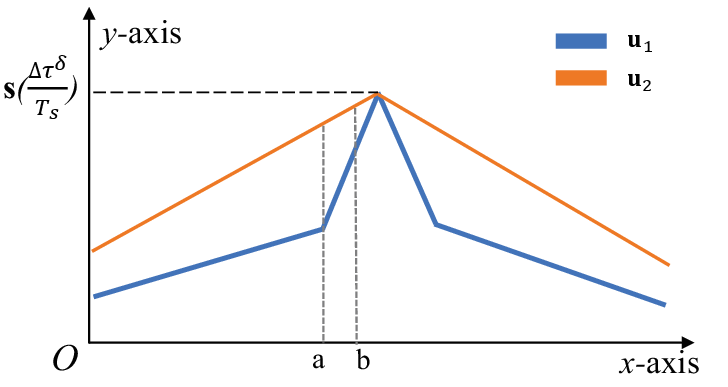}
	\caption{${\bf u}_1$ and ${\bf u}_2$ are two implementations of ${\bf s}$. ${\bf u}_1$ has a sharper mainlobe than ${\bf u}_2$.}
	\label{Pro1}
\end{figure}

In practical  high-SNR scenarios, the noise power $\bar\sigma^2$ is significantly smaller than the difference between two consecutive elements on the mainlobe of the mean sequence, ${|{\boldsymbol{\beta}_{\rm n}}|^2}{\bf s}$, such as ${|{\boldsymbol{\beta}_{\rm n}}|^2}[{\bf s}({\rm Round}(\frac{\Delta\delta^\tau}{T_{\rm R}}))-{\bf s}({\rm Round}(\frac{\Delta\delta^\tau}{T_{\rm R}}-1))]$.
In such cases, if the mainlobe becomes narrower, although $|{\boldsymbol{\beta}_{\rm n}}|^2$ may decrease, the change in $\bar\sigma^2$ is relatively smaller than the change in the difference between two adjacent elements. As a result, as shown in Fig. \ref{Pro1}, the probability that ${\bf u}_2(a)$ is higher than ${\bf u}_2(b)$ will be lower than the probability that ${\bf u}_1(a)$ is higher than ${\bf u}_1(b)$. Consequently, the maximum will be located closer to the peak point of ${\bf s}$, when ${\bf s}$ has narrower lobe. 

As (\ref{correlation}) shows, ${\bf s}$ is the linear weighted sum of the cyclic shifted $\boldsymbol{\rho}_{1,1}$.
Moreover, as (\ref{rhor1}) suggests, $\boldsymbol{\rho}_{1,1}$ can be represented as the convolution of the Fourier transforms of the windows utilized and the reversed sequences of the Fourier transforms, which is equivalent to the autocorrelation of the Fourier transform of the  window used. In accordance with the definition of the autocorrelation, it can be viewed as the sliding multiply and sum operation for a certain sequence. Then, it is intuitive that the mainlobe sharpness of the cross-correlation, namely $\boldsymbol{\rho}_{1,1}$, is related to the mainlobe sharpness of the Fourier transform of the window utilized, given the usually high attenuation of the sidelobes. 
As a result, the sharper the mainlobe of the Fourier transform of the utilized window peak is, the better the estimation performance.

\textit{Based on the above insight, we aim for finding windows whose Fourier transform has sharp mainlobe to advance the estimation performance.} However, there are no other traditional windows whose Fourier transform has narrower mainlobe width than that of the rectangular window. As a result, we find other ways and creatively utilize the super-resolution estimation algorithm (SREA) to generate suitable windows. 

Generally, we process ${\boldsymbol{\Gamma}_m}$ by replacing the traditional 2D-DFT by the SREA, which is reminiscent of the MUSIC algorithm. The outputs of the MUSIC algorithm are  bell-shaped curves centered at specific coordinates. If the MUSIC algorithm is regarded as a spectral analysis tool, just like the DFT, then the outputs of the MUSIC can be accordingly viewed as spectral components filtered by the ``bell-shaped window"\footnote{The "bell-shaped window" here refers to the shape of output of the MUSIC algorithm when a single-tone signal is input. For convenience, we name this window as the bell-shaped window with the MUSIC algorithm (BS-W-MUSIC) in what follows.}. The process of using the MUSIC algorithm for ${\boldsymbol{\Gamma}_m}$ is very similar to that of utilizing the DFT for ${\boldsymbol{\Gamma}_m}$. Thus, the ``bell-shaped window" is treated as a special window generated by the MUSIC algorithm. Given the generally sharp peak of the bell-shaped curve \cite{6942180}, the corresponding synchronization performance will be improved according to the analytical results in Section IV.

\begin{figure}[tbp]
	\centering
	\includegraphics[width=3.5in]{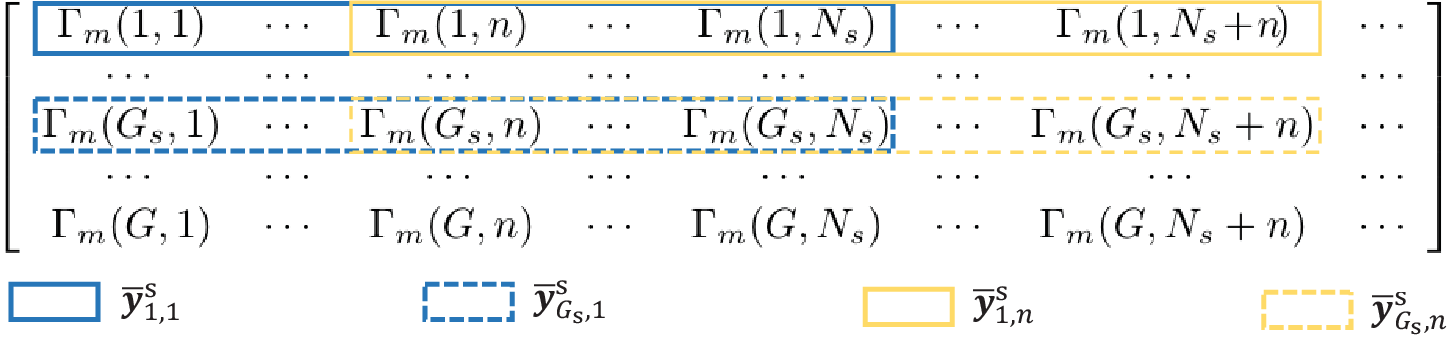}
	\caption{Frequency smoothing schemes. The matrix shown in this figure is $\boldsymbol{\Gamma}_m$. The smoothing vectors in the same color, such as $\bar {\bf y}_{1,1}^{\rm s}$ and $\bar {\bf y}_{G_s,1}^{\rm s}$, will be stacked into the same vector, $\bar {\bf y}_{1,1}$.}\label{smoothing}
\end{figure}

Concretely, to acquire a  preferable ``bell-shaped window", we first preprocess $\boldsymbol{\Gamma}_m$ by frequency-domain (FD) smoothing \cite{liu2020super,2829988} before utilizing the MUSIC algorithm. Specifically, we define $\bar {\bf y}_{g,n}^{\rm s}$ as the smoothing vector and its composition is shown in Fig. \ref{smoothing}.
Subsequently, we stack several smoothing vectors into $\bar {\bf y}_{g,n}$ as $[({\bar {\bf y}_{g,n}^{\rm s}})^{\rm T},\cdots,({\bar {\bf y}_{g+G_{\rm s},n}^{\rm s}})^{\rm T},\cdots,(\bar{\bf y}_{g+(g_{\rm s}-1)G_{\rm s},n}^{\rm s})^{\rm T}]^{\rm T}$, where $G_s$ is the index difference between two continuous vectors and $g_{\rm s}$ is the total number of  smoothing vectors stacked into $\bar {\bf y}_{g,n}$. Moreover, $N_s$ in Fig. \ref{smoothing} represents the length of the smoothing vectors.

After frequency-domain smoothing, we represent the covariance matrix 
\begin{equation}\label{newequ33}
{\bf C}=\sum\nolimits_{n=1}^{n_{\rm s}}\sum\nolimits_{g=1}^{g_{\rm s}}\bar {\bf y}_{g,n}\bar {\bf y}_{g,n}^{\rm H},
\end{equation} to implement the MUSIC algorithm.
Then, by applying the eigenvalue decomposition to ${\bf C}$, we can obtain
\begin{equation}
{\bf C}=[{\bf U_{\rm s}},{\bf U_{\rm n}}]{\boldsymbol{\Lambda}}[{\bf U_{\rm s}},{\bf U_{\rm n}}]^{\rm H},
\end{equation}
where both ${\bf U_{\rm s}}$ and ${\bf U_{\rm n}}$ are the normalized unitary matrices and ${\boldsymbol{\Psi}}$ is a diagonal matrix, whose diagonal elements are the eigenvalues of ${\bf C}$.
As a result, the metric composed by bell-shaped windows can be formulated as
\begin{equation}\label{equ18}
{\bf B}_{\rm s}(\delta,\tau)=\frac{1}{{\bf a}^{\rm H}_{\delta,\tau}{\bf U}_{\rm n}{\bf U}_{\rm n}^{\rm H}{\bf a}_{\delta,\tau}},
\end{equation}
where we have ${\bf a}_{\delta,\tau}={\boldsymbol \theta}\otimes{\boldsymbol \tau}$. Moreover, ${\boldsymbol \theta}$ and ${\boldsymbol \tau}$ are defined as $[1,\cdots,e^{-j2\pi gG_s\Delta f\tau},\cdots,e^{-j2\pi g_sG_s\Delta f\tau}]^{\rm T}$ and $[1, \cdots, e^{-j2\pi \delta(N_s-1)T_{\rm sym}}]^{\rm T}$, respectively. Since the center coordinates of the bell-shaped windows are related to $\delta$ and $\tau$, we can estimate them by locating the peaks of the metric.

Compared with the conventional usage of the MUSIC algorithm, our scheme only exploits the comparatively sharp lobe of the output of the MUSIC algorithm to generate a superior fingerprint spectrum derived from BS-W-MUSIC, rather than directly estimating the Doppler frequency shift and the time delay \cite{liu2020super}. Then, by leveraging the fingerprint spectrum, the cross correlation in (\ref{unsimplified}) is harnessed for estimating the CFO and TO. However, the conventional MUSIC algorithm is not directly applicable to CFO and TO estimation due to the associated clock asynchronism issue.



\section{Numerical Simulations}

\begin{figure}
	\centering
	\hspace*{-0.6cm}
	\includegraphics[width=3.5in]{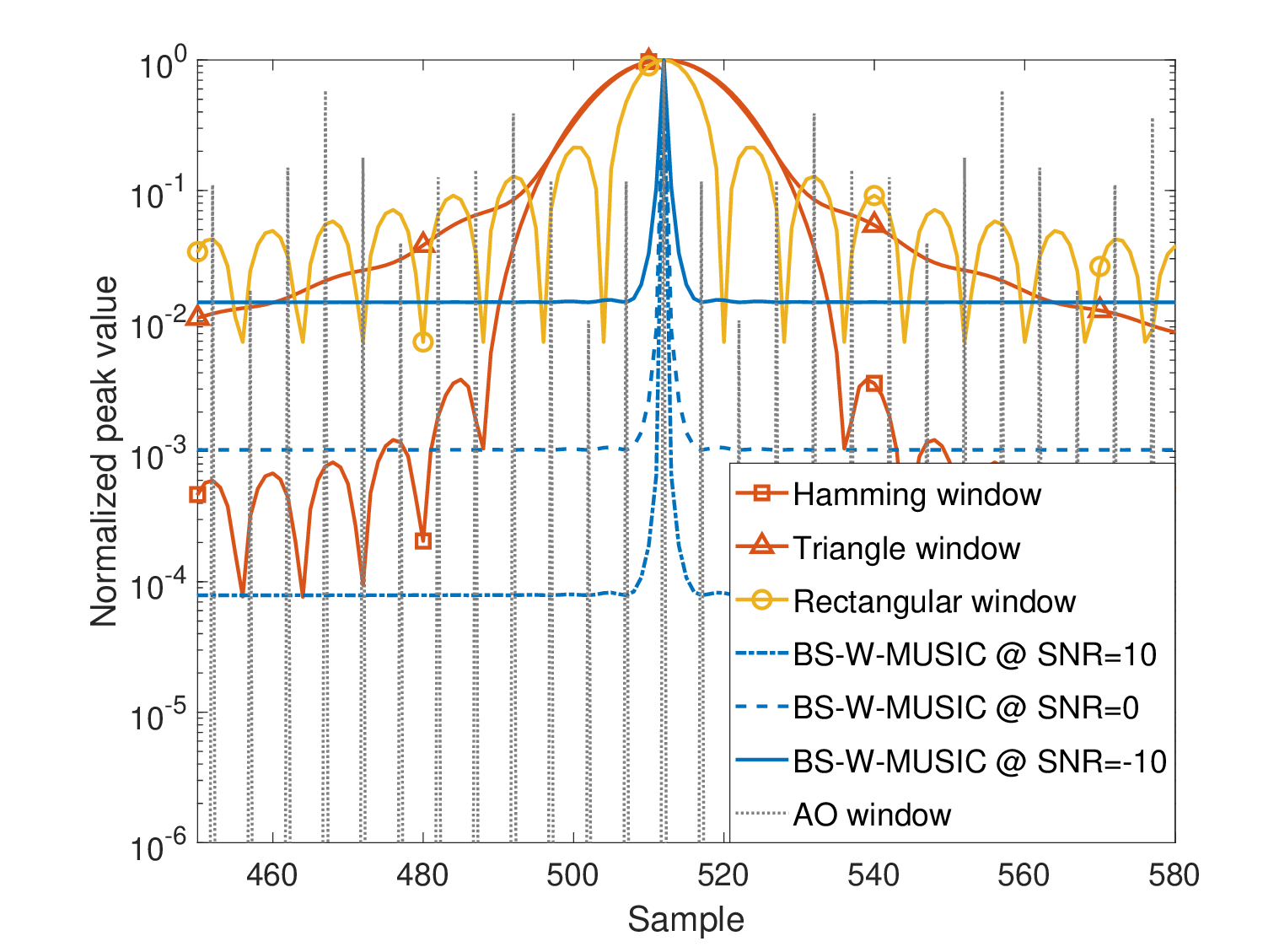}
	\caption{Cross-correlation of the BS-W-MUSIC, Fourier transforms corresponding to different traditional window functions and the asymptotically optimal window. 
	``AO window" represents correlation of our proposed asymptotically optimal window. }\label{Correlation}
\end{figure}

In this section we perform numerical simulations for verifying the accuracy of our theoretical analysis and for evaluating the synchronization performance, when employing a bell-shaped window. In the simulations, we assume that the number of RRU and UE antennas is $M_{\rm r}=64$ and $M_{\rm t}=2$, respectively. The carrier frequency $f_c$ of this system  is assumed to be $28$ GHz and the subcarrier spacing is set to $100$ KHz, which is similar to \cite{ni2021uplink,rahman2019framework}. Moreover, we assume that the number of subcarriers is $128$ and the length of CP is $16$. Based on these parameters, the sampling interval can be simply calculated as $7.8125\times 10^{-8}$ s and the length of the whole OFDM symbol is  $11.25$ $\mu$s. The time length suggests that even though several hundreds of OFDM symbols are employed for sensing, it will take only a few milliseconds, corresponding to near-real-time perception. 

\begin{figure}
	\centering
	\includegraphics[width=3.5in]{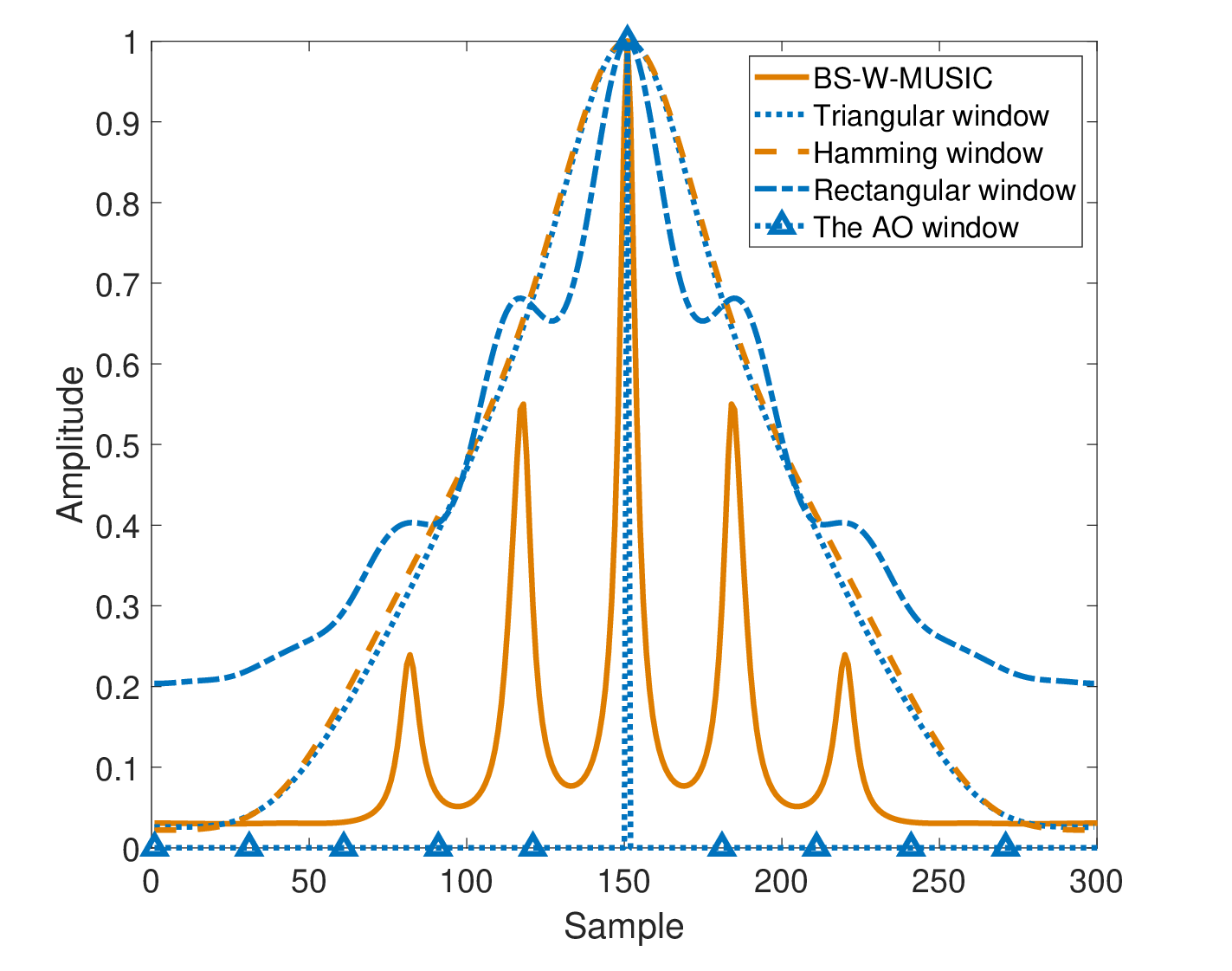}
	\caption{The auto-correlation sequence of fingerprint spectra corresponding to different windows. The fingerprint spectra are generated by randomly setting 4 static objects in the environmental background. Moreover, the received signals reflected from different objects are randomly set as $[10, 4, 7, 1]$ dB and the propagation distances of these paths are randomly set as $[40, 50, 75, 110]$ meters. Moreover, the ``The AO window" represents the auto-correlation sequence of fingerprint spectrum generated by the desired asymptotically optimal window. 
	}\label{Correlation_fingerprint}
\end{figure}

\begin{figure}
	\centering
	\subfigure[The 3D spectrum corresponding to the rectangular window.]{
		\begin{minipage}[t]{1\linewidth}
			\centering
			\includegraphics[width=3.5in]{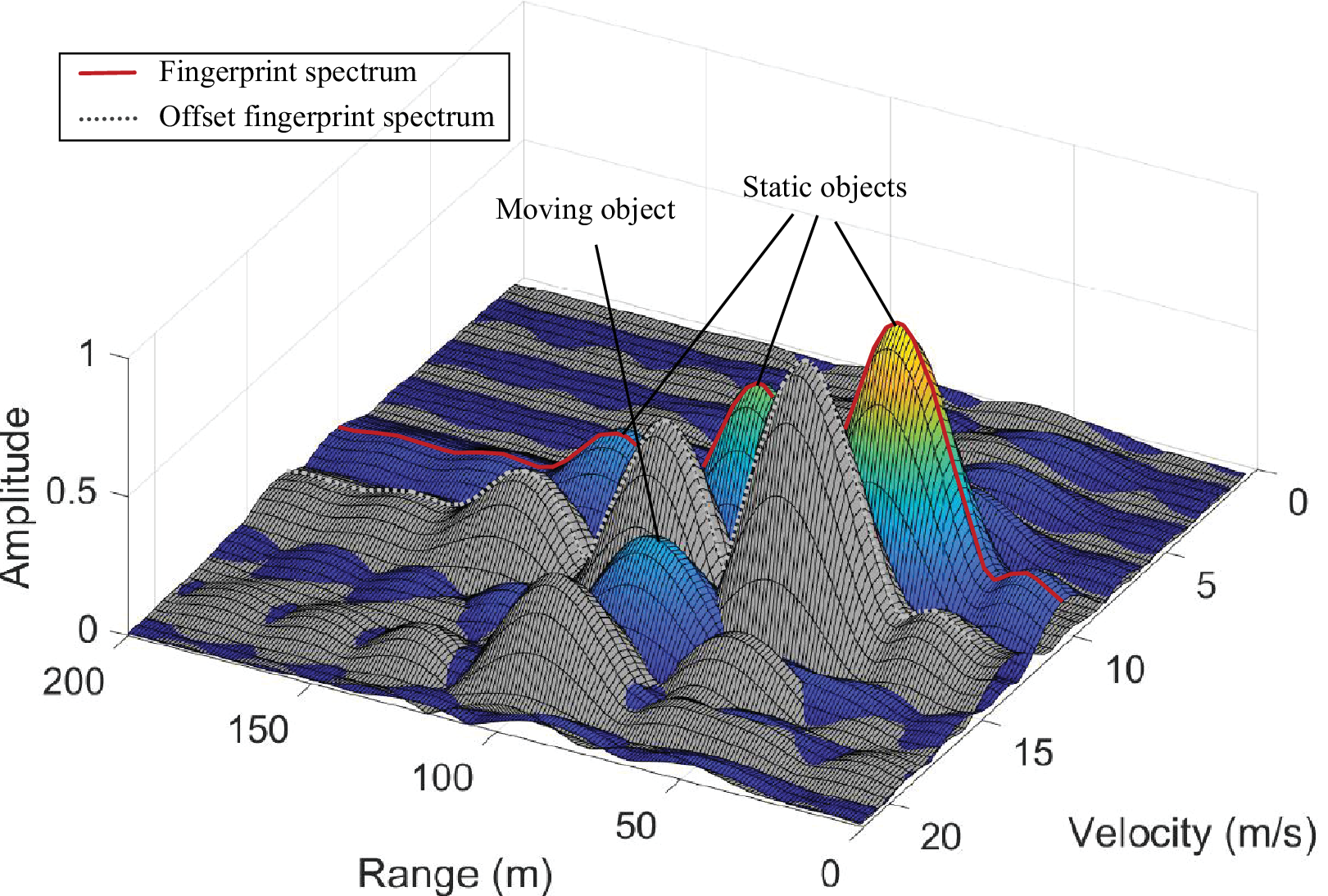}\\
		\end{minipage}%
	}%
	
	\subfigure[The 3D spectrum corresponding to the proposed BS-W-MUSIC.]{
		\begin{minipage}[t]{1\linewidth}
			\centering
			\includegraphics[width=3.5in]{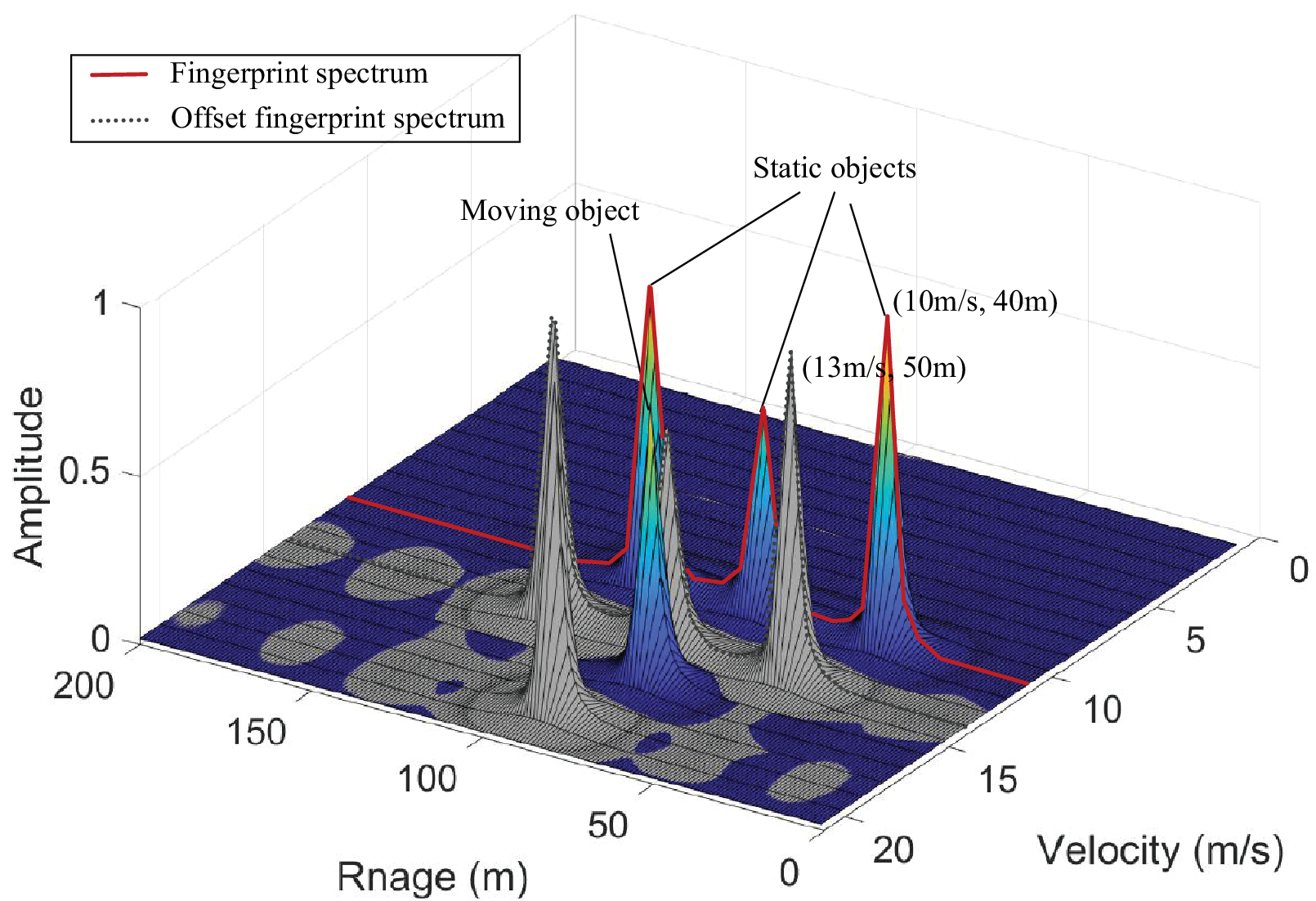}\\
		\end{minipage}%
	}%
	\centering
	\caption{The 3D spectrum implementations for multiple windows. The spectra are  generated by randomly setting several static objects and dynamic object in the environmental with/without CFO and TO. Moreover, the spectrum corresponding to the rectangular window is generated by utilizing 25-fold zero-padding along the Range-axis and 5-fold zero-padding along the Velocity-axis. The blue 3D spectrum and the grey one represent the initial spectrum and the offset spectrum with CFO and TO drift. The set CFO and TO are equivalent to the Doppler frequency shift induced by a small velocity of $3~m/s$, and the time delay resulting from a small distance of $10~m$, respectively. The red outlines shown in both sub-figures are selected as the fingerprint spectrum sequence, while the dashed grey lines represent the offset fingerprint spectrum sequence.}
	\label{fig7}
\end{figure}

In Fig. \ref{Correlation} we draw the auto-correlation sequence of the Fourier transforms of different windows, including the traditional windows, the bell-shaped window and the near-optimal window. Specifically, the length of the  traditional windows is set to $128$, which equals to the number of subcarriers. Moreover, a 25 times zero-padding is utilized, when applying the DFT to these windows, which means that $128\times24$ zeros will be appended to the windows before applying the DFT. For intuitive comparison, the bell-shaped window is generated by applying the MUSIC algorithm to a $128$-subcarrier system. Furthermore, since the shape of peaks output by MUSIC is influenced by several factors, including the SNR and the number of antennas as well as snapshots \cite{6942180}, we evaluate the lobe width of bell-shaped window for diverse parameter combinations. Although the noise floor of the bell-shaped window varies when SNR and the number of snapshots increase, the width of the mainlobe of the window is invariably lower than that of all the traditional windows. Moreover, since the asymptotically optimal window shape is dependent on a specific scenario, we draw its correlation in the same scenarios as those in Fig. \ref{Correlation_fingerprint} for intuitive comparison. Observing in Figure \ref{Correlation} that both the mainlobe and the sidelobes associated with the 'AO window' are remarkably sharp, and the sidelobes are more prominent compared with those of the BS-W-MUSIC and conventional windows. Nevertheless, the amplitude of each sidelobe and the spacing between them in the 'AO window' have been carefully designed based on the gains and time delays of each path, as described in Proposition \ref{pro2}. This design allows for the cancellation of sidelobes, while preserving the mainlobe during the generation of the fingerprint spectrum (which involves the summation of multiple cyclic-shifted AO windows). As a result, despite the initially large sidelobes in the AO window, they disappear in the fingerprint spectrum. Conversely, for the BS-W-MUSIC and conventional windows, which have wider sidelobes, sidelobe cancellation can be hardly achieved. In such cases, the windows having sharper sidelobes may yield a fingerprint spectrum associated with narrower sidelobes, thereby leading to improved synchronization performance.

In Fig. \ref{Correlation_fingerprint}, we draw the cross-correlation sequence of the fingerprint spectra generated by the traditional windows, bell-shaped windows and the asymptotically optimal window. As the figure shows, the fingerprint spectrum generated by the bell-shaped window exhibits sharper peaks than the mainlobe of the traditional windows,  while the peak of the asymptotically optimal window is the sharpest. Actually, the shape of ``AO window" curve is formulated as (\ref{s_q}). Moreover, to intuitively observe the shape of these fingerprint spectra, we also draw the 3D spectra shown in Fig. \ref{fig7},  and the fingerprint spectra utilized to draw Fig. \ref{Correlation_fingerprint} is actually selected from the row marked as red in Fig. \ref{fig7}. For brevity, only the fingerprint spectrum generated by the rectangular window is drawn in Fig. \ref{fig7}, due to its narrowest mainlobe among all the traditional windows. By observing all the different spectra in Fig. \ref{fig7}, it becomes evident that according to our analysis, the synchronization performance of the BS-W-MUSIC is better than that of traditional windows. Moreover, according to Fig. \ref{fig7}, the impact of CFO and TO on the fingerprint spectrum can be intuitively observed.


\begin{figure}
	\centering
	\includegraphics[width=3.5in]{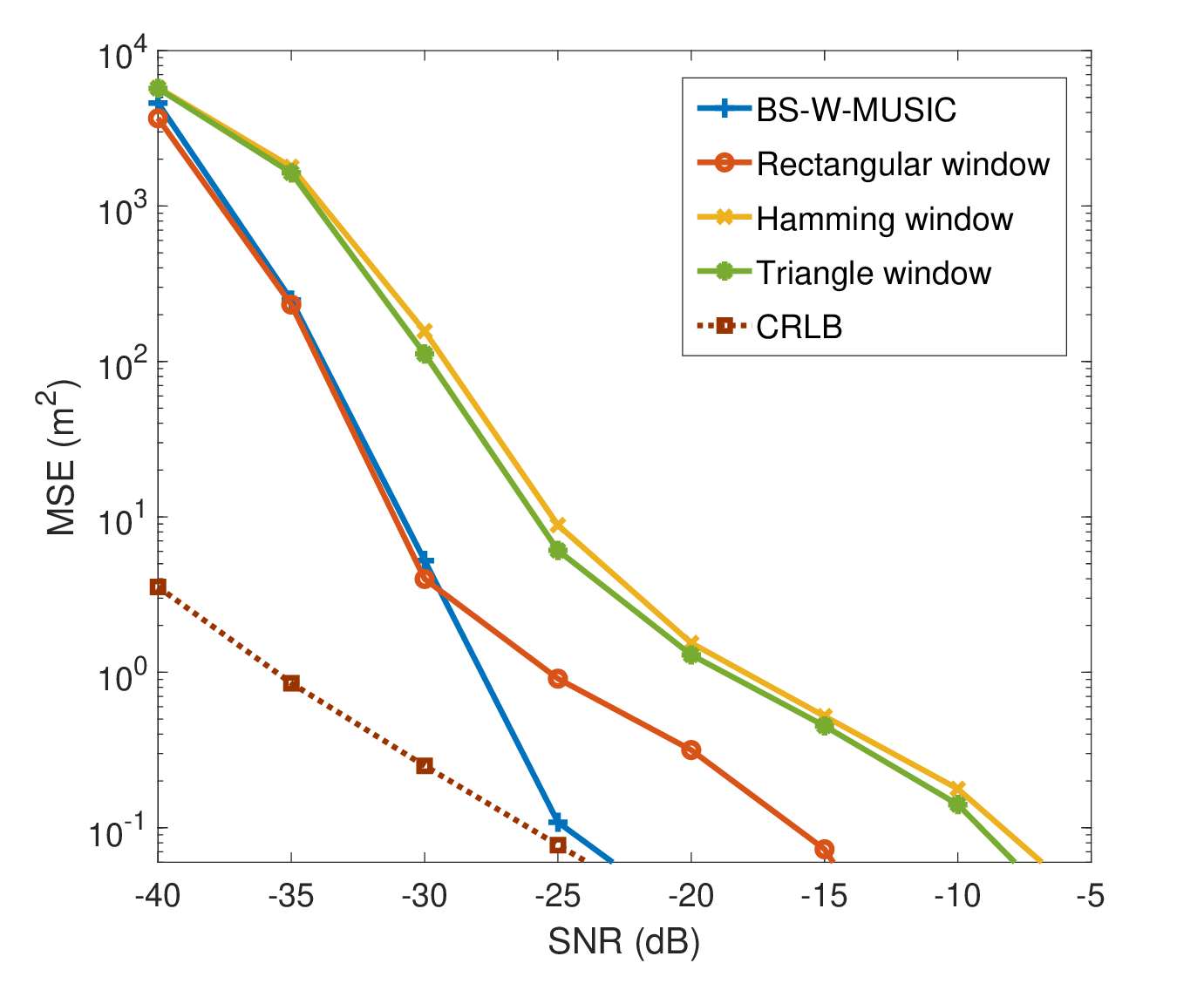}
	\caption{The synchronization MSE of different windows and the BS-W-MUSIC. 
	}\label{fig10}
\end{figure}

\begin{figure}[htb]
	\centering
	\includegraphics[width=3.5in]{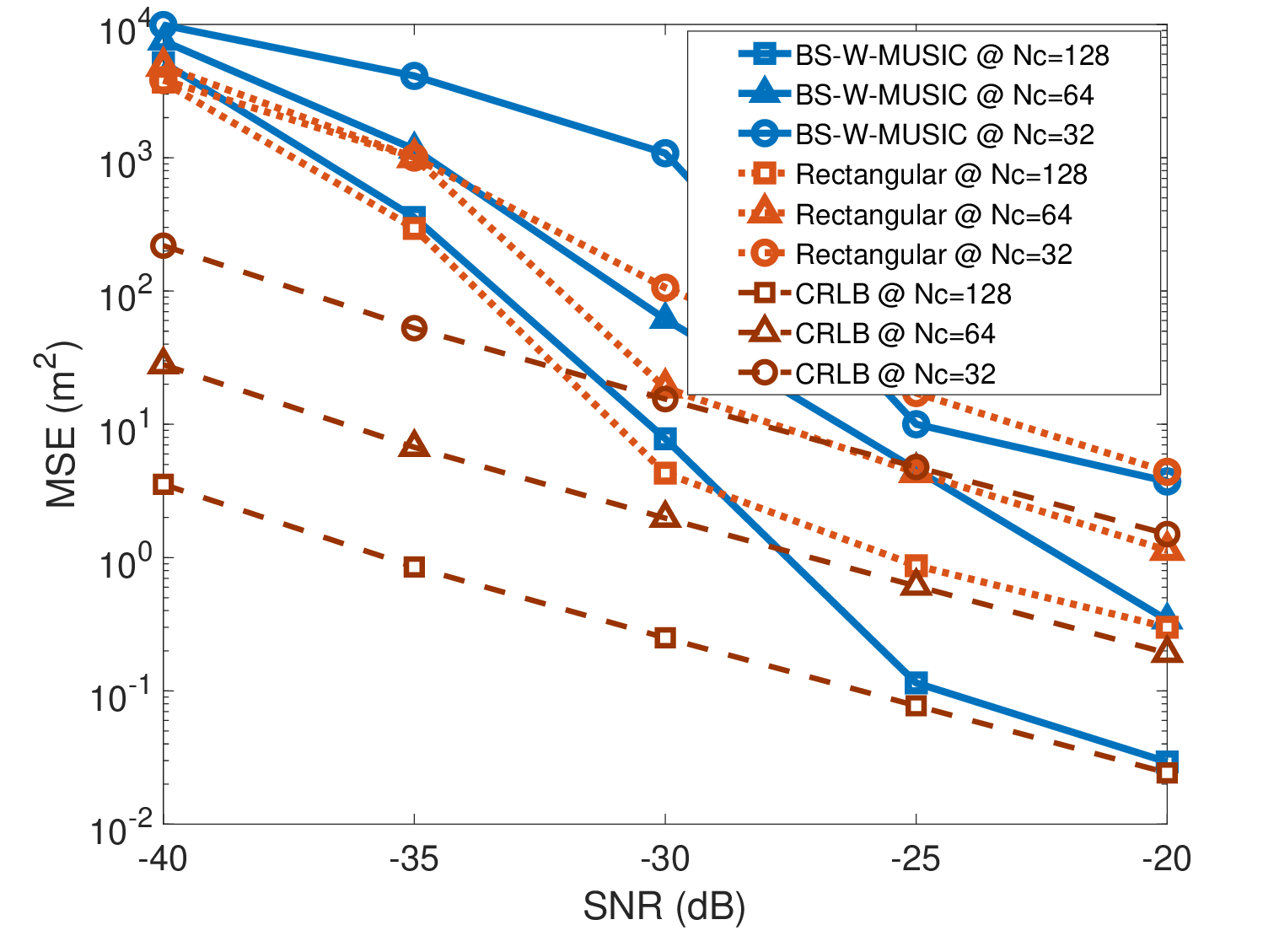}
	\caption{The synchronization MSE of BS-W-MUSIC and the rectangular window while setting different number of subcarriers.}\label{MSE_windows-music-4}
\end{figure}

\begin{figure}[htb]
	\centering
	\includegraphics[width=3.5in]{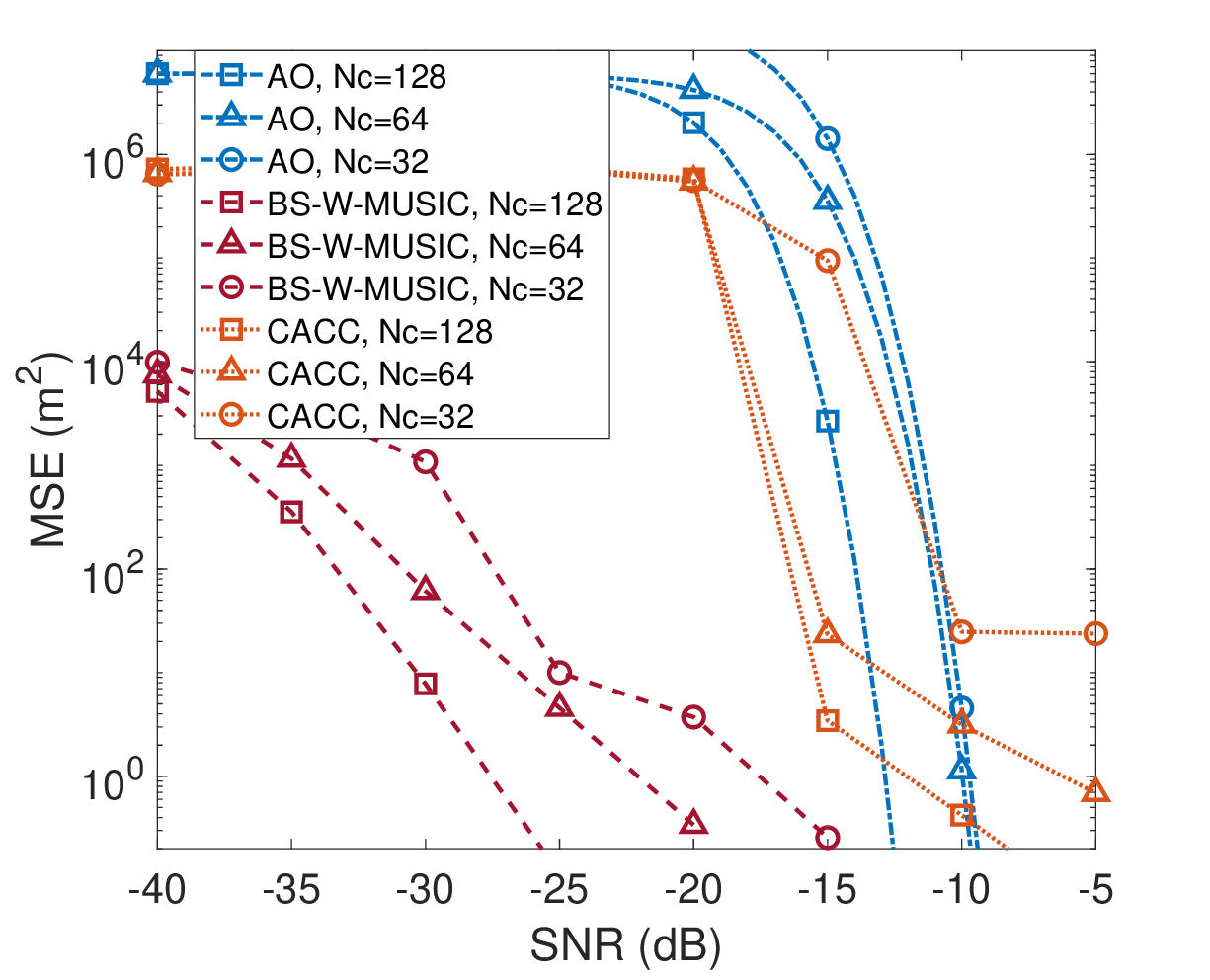}
	\caption{The estimation MSE of the asymptotically optimal window, the BS-W-MUSIC, and the CACC synchronization method.
	}\label{fig11}
\end{figure}

In Fig. \ref{fig10}, we evaluate the synchronization performance of different windows by locating the peaks of the autocorrelation sequence of the corresponding fingerprint spectrum using Monte Carlo simulations. We run thousands of Monte Carlo estimation trials to generate the synchronization MSE for different SNR, number and locations of static objects. Concretely, the average MSE is defined as
\begin{equation}
{\rm MSE} = \frac{1}{N_{\rm simu}}\sum_{n=1}^{N_{\rm simu}}[R\times(\hat{l}-l_{\rm R})]^2,
\end{equation}
where  $l_{\rm R}$ and $\hat{l}$ are the ground truth and the estimated indices by using (\ref{unsimplified}), respectively. Moreover, $R$ is the propagation distance corresponding to the unit coordinate  and $N_{\rm simu}$ represents the number of numerical simulation runs. In this simulation, we employ $400$ OFDM symbols each time for sensing and these symbols will approximately have a length of 4.5 ms in practical scenarios. This short time ensures the timeliness of perception. Additionally, to diversify the parameter configuration, we randomly select the number of static objects in the environment from the set of $\{2,3,4\}$. It is important to note that the vertical axis unit used to evaluate the peak estimation performance of the autocorrelation sequence of the fingerprint spectrum is square meters (m²). This choice is made because calculating the MSE as outlined in (\ref{theoretical_MSE}), based on the MSE of the range ambiguity induced by the TO estimation error, offers a more intuitive measure of the performance. Moreover, this unit applies to both Fig. \ref{MSE_windows-music-4} and Fig. \ref{fig11}.

As shown in Fig. \ref{fig10}, it is evident that the rectangular window has the best performance among the traditional windows, while the bell-shaped window outperforms the rectangular window. This phenomenon conforms to our insight. However, the performance of the bell-shaped window is better than that of the rectangular window, when the SNR is higher than $-30$ dB. This phenomenon appears, since  the noise floor of the bell-shaped window increases with the SNR decreasing and consequently the peak location performance becomes worse than that of the rectangular window. Moreover, there is rapid drop in the curve for the proposed BS-W-MUSIC, when the SNR is in the range of $(-40,-30)$ dB. The drop emerges since the peak changes from inconspicuous at $-40$ dB to obvious at $-30$ dB, and that change makes the peak location more accurate. However, when the SNR becomes higher, the increase in SNR results in a diminishing  enhancement of the positioning accuracy.

In Fig. \ref{MSE_windows-music-4}, we compare the synchronization performance of the rectangular window and of the BS-W-MUSIC for different numbers of subcarriers. A pair of insights may be drawn from Fig. \ref{MSE_windows-music-4}. The first is: both the performance of the rectangular window and of the bell-shaped window gets better with the number of subcarriers increasing, which is intuitive. As the CRLB we derived in \cite{wxy} suggests: the larger the number of subcarriers, the lower the CRLB. Concretely, the CRLB matrix is formulted as ${\rm CRLB} = \sigma^2\left(\sum_{n=1}^{N_{\rm c}} {\bf H}_n^{\rm H}{\bf H}_n\right)^{-1}$,
where ${\bf H}_n=[\frac{\partial\boldsymbol{\Gamma}_m}{\partial v_{1}},\cdots,\frac{\partial\boldsymbol{\Gamma}_m}{\partial v_{l}},\frac{\partial\boldsymbol{\Gamma}_m}{\partial \tau_{1}},\cdots,\frac{\partial\boldsymbol{\Gamma}_m}{\partial \tau_{l}}]$.
The second insight is: regardless of the specific subcarrier configuration, the performance of the bell-shaped window is better than that of the rectangular window, when the SNR achieves critical point, like $-35$dB for $N_c=128$. However, the point will generally move to the left with the number of subcarriers increasing, because the MUSIC algorithms performs better, when the length of the steering vector is higher. 

In Fig. \ref{fig11}, we evaluate the theoretical MSE of the asymptotically optimal window and compare it with that of the BS-W-MUSIC and the CACC synchronization method using numerical simulations, respectively. 
As Fig. \ref{fig11} shows, for all the three methods, the MSE improves as the number of subcarriers increases. Moreover, the MSE drops rapidly when the SNR exceeds $-10$ dB for the asymptotically optimal window. However, under low-SNR conditions, the MSE of peak location degrades severely despite the impulse function's excellent peak location performance under high-SNR scenarios. Interestingly, the BS-W-MUSIC and CACC methods outperform the method of using the asymptotically optimal window, suggesting that a narrower mainlobe is not always better in low-SNR scenarios and wider mainlobe windows may perform better. Furthermore, the MSE of our BS-W-MUSIC method is lower than that of the CACC method. When considering Fig. 9 alongside this observation, it becomes apparent that the MSE performance of CACC is roughly equivalent to that of using the Hamming window. Another notable observation in Fig. 11 is that the rate of MSE decrease associated with the asymptotically optimal window is significantly faster than that of the BS-W-MUSIC and the CACC methods. This is due to the design of the asymptotically optimal window relying on the exact gains and delays of each propagation path, allowing for more known information to estimate the CFO and TO than when using traditional windows. 
However, it is important to note that it is infeasible to achieve the theoretical lower bound of the MSE for the asymptotically optimal window in practical systems.

\section{Conclusions}
Window functions  designed have been investigated for fingerprint-spectrum-based passive sensing tailored to PMNs. Concretely, we derived the expressions of an  asymptotically optimal window and of the corresponding synchronization MSE. However, the asymptotically optimal window cannot be practically harnessed since it is strongly dependent on the specific time delays and gains corresponding to each propagation path, which are unknown before successful sensing. 
To circumvent this issue, we tested a practical bell-shaped window relying on super-resolution estimation algorithms and simulations have verified our theoretical analysis. To some degree, this work reveals the relationship between the synchronization performance and the  window function utilized in the PMN systems, where sensing is achieved by spectrum analysis. Future improvements of this work may be conducted by finding the exact globally optimal window or proposing a closed-form metric to accurately evaluate the synchronization performance of hundreds of windows.

\appendices
\section{Proof of proposition 1}\label{proof1}
According to (\ref{theoretical_MSE}), to minimize $f_{\rm MSE}(\boldsymbol{\psi}_{N_c}) $, the decision variables, namely ${\rm P}_{\{q,KN_c\}}$ for $q=1,\cdots,KN_c$, should be optimized. Based on the definition of ${\rm P}_{\{q,KN_c\}}$, namely (\ref{newequ15}), ${\rm P}_{\{q,KN_c\}}$ is only determined by the fingerprint spectrum, namely the $K_1$th row of ${\bf A}[K_1,q]$.  Moreover,
${\bf A}[K_1,q]$ can be deemed to be a complex Gaussian variable with a mean of $\sum_{p=1}^{KN_c}{\boldsymbol{\beta}}[p]{\boldsymbol{\beta}}_{\rm c}[p+q]$ for $q=1,\cdots,KN_c$ and variance of (refer to Appendix \ref{noise})
\begin{equation}\label{sigma}
\bar\sigma^2\approx\frac{2{\tilde\sigma^2}}{{|{\boldsymbol{\beta}_{\rm n}}^*|^2}}+\frac{KN_c}{2{|{\boldsymbol{\beta}_{\rm n}}^*|^4}}{\tilde\sigma^2}.
\end{equation}
Equation (\ref{sigma}) suggests that if the noise power of the received signal is given, $\tilde\sigma^2$ will be fixed. Consequently, $\bar\sigma^2$ will be determined by the power of the fingerprint spectrum, $|{\boldsymbol{\beta}_{\rm n}}^*|^2$. Moreover, since the maximum value of ${\bf s}$ is normalized to 1, $|{\boldsymbol{\beta}_{\rm n}}^*|^2$ approximately equals the maximum element of $|{\boldsymbol{\beta}_{\rm n}}^*|^2{\bf s}$. Thus, the minimized MSE can be determined by finding the optimal ${\bf s}$. 
Specifically, we determine the optimal ${\bf s}$ in two steps.

\textit{Step 1}: By assuming that  $\frac{\Delta\delta^\tau}{T_{\rm R}}$ is an integer, we first clarify that for any value of ${\bf s}(q)$ for $q\neq\frac{\Delta\delta^\tau}{T_{\rm R}}$, $f_{\rm MSE}(\boldsymbol{\psi}_{N_c})$ is minimized when $|{\boldsymbol{\beta}_{\rm n}}^*|^2{\bf s}[\frac{\Delta\delta^\tau}{T_{\rm R}}]$ is maximized. Note that ${\bf s}[\frac{\Delta\delta^\tau}{T_{\rm R}}]=1$ in this case, according to the definition of ${\bf s}$. 

\textit{Step 1.1}: According to (\ref{sigma}), the normalized noise power $\bar{\sigma}^2$ decreases upon increasing $|{\boldsymbol{\beta}_{\rm n}}^*|^2{\bf s}[\frac{\Delta\delta^\tau}{T_{\rm R}}]$  for a fixed $\tilde\sigma^2$. Then, the probability ${\rm P}_{\{q,KN_c\}}$ for $q\neq\frac{\Delta\delta^\tau}{T_{\rm R}}$ will decrease, while ${\rm P}_{\{{\Delta\delta^\tau}/{T_{\rm R}},KN_c\}}$ will increase. To prove this, we represent ${\rm P}_{\{q,KN_c\}}$ as 
\begin{small}
\begin{equation}\label{P_CDF1}
\begin{aligned}
&{\rm P}_{\{q,KN_c\}}=\int_{V}p({\bf s},\tilde\sigma^2){\mathrm d}\ {\bf b}\\
&=\!\!\int_{-\infty}^{\infty}\!\!\frac{e^{-{[{ b}_q-{\bf s}[q]]^2}/{2\tilde\sigma^2}}}{\sqrt{2\pi\tilde\sigma}}\!\!\left(\prod \limits_{i=1,i\neq q}^{KN_c}\int_{-\infty}^{b_q}\frac{e^{-{ [{ b}_i-{\bf s}[i]]^2}/{2\tilde\sigma^2}}}{\sqrt{2\pi\tilde\sigma}}{\mathrm d}b_i\right)\!\!{\mathrm d}b_q\\
&=\!\!\int_{-\infty}^{\infty}f_{\mathcal N}(b_q)\prod \limits_{i=1,i\neq q}^{KN_c} Q(\frac {b_q-{\bf s}[i]}{\bar\sigma^2}){\mathrm d}\ b_q,
\end{aligned}
\end{equation}\end{small}where $f_{\mathcal N}(\cdot)$ represents the probability density function of the standard normal distribution. From the formulation, it is observed that ${\rm P}_{\{q,KN_c\}}$ could be expressed as the integral of $\prod \limits_{i=1,i\neq q}^{KN_c} Q(\frac {b_q-{\bf s}[i]}{\bar\sigma^2})$. For any value of ${\bf s}[q],\ q\neq\frac{\Delta\delta^\tau}{T_{\rm R}}$, $\prod \limits_{i=1,i\neq q}^{KN_c} Q(\frac {b_q-{{\bf s}[i]}}{\bar\sigma^2})$ will reduce when $|{\boldsymbol{\beta}_{\rm n}}^*|^2{{\bf s}[\frac{\Delta\delta^\tau}{T_{\rm R}}]}$ increases and as a result, ${\rm P}_{\{q,KN_c\}}$ reduces. By contrast, for $q=\frac{\Delta\delta^\tau}{T_{\rm R}}$, $\prod \limits_{i=1,i\neq q}^{KN_c} Q(\frac {b_q-{{\bf s}[i]}}{\bar\sigma^2})$ increases when $|{\boldsymbol{\beta}_{\rm n}}^*|^2{{\bf s}[\frac{\Delta\delta^\tau}{T_{\rm R}}]}$ increases and consequently ${\rm P}_{\{{\Delta\delta^\tau}/{T_{\rm R}},KN_c\}}$ increases. Thus, $|{\boldsymbol{\beta}_{\rm n}}^*|^2{{\bf s}[\frac{\Delta\delta^\tau}{T_{\rm R}}]}$ should be set as large as possible to minimize $f_{\rm MSE}(\boldsymbol{\psi}_{N_c})$.

\textit{Step 1.2}: Upon confirming the optimal value of $|{\boldsymbol{\beta}_{\rm n}}^*|^2{\bf s}[\frac{\Delta\delta^\tau}{T_{\rm R}}]$, we then determine the optimal ${\bf s}[q],\ q\neq\frac{\Delta\delta^\tau}{T_{\rm R}}$. Observe from (\ref{theoretical_MSE}),  that $f_{\rm MSE}(\boldsymbol{\psi}_{N_c}) $ can be deemed to be a linear weighted sum of elements in $\{(1-\frac{\delta^\tau}{T_{\rm R}})^2,\cdots,(KN_c-\frac{\delta^\tau}{T_{\rm R}})^2\}$ and the weights are $\{{\rm P}_{\{1,KN_c\}},\cdots,{\rm P}_{\{KN_c,KN_c\}}\}$. Moreover, it also holds that $\sum_{q=1}^{KN_c}{\rm P}_{\{q,KN_c\}}=1$. 
As a result, to minimize $f_{\rm MSE}(\boldsymbol{\psi}_{N_c})$, we should assign as much weight as possible to the index closest to $\frac{\Delta\delta^\tau}{T_{\rm R}}$ and as low weight as possible to indices far from $\frac{\Delta\delta^\tau}{T_{\rm R}}$. However, it is difficult to design the optimal ${\bf s}$ to minimize $f_{\rm MSE}(\boldsymbol{\psi}_{N_c})$ utilizing the complex integral of the $Q$ function. 


\textit{Step 2}:
Consequently, to find superior window functions, we aim for finding an approximate optimal solution, when the SNR tends to infinity. Specifically, we define ${\bf s}_{\rm op}$ as the undetermined optimal fingerprint spectrum and denote the approximate optimal fingerprint spectrum by ${\bf s}_{\rm ap}$. In the following, we prove that \textit{when ${\bf s}_{\rm ap}$ satisfies}
\begin{equation}\label{ideal1}
\begin{cases}
{\bf s}_{\rm ap}[q]=1, q=\frac{\Delta\delta^\tau}{T_{\rm R}},\\
{\bf s}_{\rm ap}[q]=0, q\neq\frac{\Delta\delta^\tau}{T_{\rm R}},
\end{cases}
\end{equation} 
\textit{the MSE tends to approach that of the optimal fingerprint spectrum.} Let us denote the probability corresponding to ${\bf s}_{\rm ap}$ and ${\bf s}_{\rm op}$ by ${\rm P}_{\{q,KN_c\}}^{\rm op}$ and ${\rm P}_{\{q,KN_c\}}^{\rm ap}$, respectively. Then, we have
\begin{equation}
\begin{aligned}
&f_{\rm MSE}({\bf s}_{\rm ap})\!\!-\!\!f_{\rm MSE}({\bf s}_{\rm op})\!\!=\!\!\sum_{q=1}^{KN_c}\!\!\left[{\rm P}^{\rm ap}_{\{q,KN_c\}}\!\!-\!\!{\rm P}^{\rm op}_{\{q,KN_c\}}\right]\!\!(q\!-\!\frac{\Delta_\delta}{T_{\rm R}})^2,
\end{aligned}
\end{equation}
where ${\rm P}^{\rm ap}_{\{q,KN_c\}}\!\!-\!\!{\rm P}^{\rm op}_{\{q,KN_c\}}$ can be represented as
\begin{small}
	\begin{equation}
	\begin{aligned}
	&{\rm P}_{\{q,KN_c\}}^{\rm ap}-{\rm P}_{\{q,KN_c\}}^{\rm op}=\\
	&\!\int_{-\infty}^{\infty}\!\!f_{\mathcal N}(b_q)\!\Bigg[\!\!\prod \limits_{i=1,i\neq q}^{KN_c} Q(\frac {b_q-{\bf s}_{\rm ap}(i)}{\bar\sigma^2})\!\!-\!\!\!\!\prod \limits_{i=1,i\neq q}^{KN_c}\!\! Q(\frac {b_q-{\bf s}_{\rm op}(i)}{\bar\sigma^2})\!\Bigg]{\mathrm d}b_q.
	\end{aligned}
	\end{equation}\end{small}With the SNR of the PMN system, namely $1/\tilde{\sigma}^2$, increasing, both $Q(\frac {b_q-{\bf s}_{\rm ap}(i)}{\bar\sigma^2})$ and $Q(\frac {b_q-{\bf s}_{\rm op}(i)}{\bar\sigma^2})$ decrease. Moreover, given any positive number $\epsilon$, there exists an SNR, defined as $1/\tilde{\sigma}^2_{\epsilon }$, which satisfies 
\begin{small}
\begin{equation}\label{con}
\Big|\prod \limits_{i=1,i\neq q}^{KN_c} Q(\frac {b_q-{\bf s}_{\rm ap}(i)}{\bar\sigma^2})\!-\!\!\prod \limits_{i=1,i\neq q}^{KN_c}\!\! Q(\frac {b_q-{\bf s}_{\rm op}(i)}{\bar\sigma^2})\Big|<\epsilon,
\end{equation}\end{small}for any $q=1,\cdots,KN_c$.

Thus, ${\bf s}_{\rm ap}$ represented as (\ref{ideal1}) is the near-optimal solution, when ${\Delta\delta^\tau}/{T_{\rm R}}$ is an integer. Moreover, to characterize the synchronization MSE in this case, we can express ${\rm P}_{\{q,KN_c\}}$ for $q\in\{1,\cdots,KN_c\}$ and $q\neq {\Delta\delta^\tau}/{T_{\rm R}}$ as
\begin{equation}
{\rm P}_{\{q,KN_c\}}=\int_{-\infty}^{\infty}\!\!f_{\mathcal N}(b_q)Q(\frac{b_q-1}{\bar\sigma^2})Q^{(KN_c-2)}(\!{b_q}\!/\!{\bar\sigma^2}\!){\mathrm d} b_q.
\end{equation}
Furthermore, the MSE in this case can be formulated as
\begin{small}
\begin{equation}
\begin{aligned}
f_{\rm MSE}(\boldsymbol{\psi}_{N_c})=&\sum_{\substack{q=1,\\ q\neq {\Delta\delta^\tau}/{T_{\rm R}}}}^{KN_c}\left[q_{\rm r}-\frac{\Delta\delta^\tau}{T_{\rm R}}\right]^2\int_{-\infty}^{\infty}\left[f_{\mathcal N}(b_q)\right.\\
&\left.Q(\frac{b_q-1}{\bar\sigma^2})Q^{(KN_c-2)}(\!{b_q}\!/\!{\bar\sigma^2}\!){\mathrm d}\right] b_q.
\end{aligned}
\end{equation}\end{small}However, since the $Q(\cdot)$ function is not in closed form, the infimum cannot be transformed into closed form.

\section{Proof of the noise $\bar\sigma^2$}\label{noise}
According to (\ref{unsimplified}), ${\bf A}[K_1,q]$ can be represented as
\begin{small}
\begin{equation}
\begin{aligned}
&{\bf A}[K_1,q]=\frac{1}{{|{\boldsymbol{\beta}_{\rm n}}^*|^2}}\sum_{p=1}^{KN_c}({\boldsymbol{\beta}}[p]+\tilde{\bf W}_K[K_0,p])({\boldsymbol{\beta}}_{\rm c}[p+q]\\
&+\tilde{\bf W}_K[K_0,p+q])=\frac{1}{{|{\boldsymbol{\beta}_{\rm n}}^*|^2}}\sum_{p=1}^{KN_c}({\boldsymbol{\beta}}[p]{\boldsymbol{\beta}}_{\rm c}[p+q]\\
&+{\boldsymbol{\beta}}_{\rm c}[p+q]\tilde{\bf W}_K[K_0,p]+{\boldsymbol{\beta}}[p]\tilde{\bf W}_K[K_0,p+q]\\
&+\tilde{\bf W}_K[K_0,p]\tilde{\bf W}_K[K_0,p+q]).
\end{aligned}
\end{equation}\end{small} It is plansible that the terms $\sum_{p=1}^{KN_c}{\boldsymbol{\beta}}[p]{\boldsymbol{\beta}}_{\rm c}[p+q]$, $\sum_{p=1}^{KN_c}{\boldsymbol{\beta}}_{\rm c}[p+q]\tilde{\bf W}_K[K_0,p]$, $\sum_{p=1}^{KN_c}{\boldsymbol{\beta}}[p]\tilde{\bf W}_K[K_0,p+q]$, and $\sum_{p=1}^{KN_c}\tilde{\bf W}_K[K_0,p]\tilde{\bf W}_K[K_0,p+q]$ are all independent. As a result, ${\bf A}[K_1,q]$ can be regarded as following a Gaussian distribution having the mean of
\begin{small}\begin{equation}
{\mathrm E}({\bf A}[K_1,q])=\frac{1}{{|{\boldsymbol{\beta}_{\rm n}}^*|^2}}\sum_{p=1}^{KN_c}{\boldsymbol{\beta}}[p]{\boldsymbol{\beta}}_{\rm c}[p+q].
\end{equation}\end{small}Moreover, the variance of the distribution is
\begin{footnotesize}
\begin{equation}
\begin{aligned}
&{\mathrm D}({\bf A}[K_1,q])=\frac{1}{{|{\boldsymbol{\beta}_{\rm n}}^*|^4}}\left[{\mathrm D}(\sum_{p=1}^{KN_c}{\boldsymbol{\beta}}_{\rm c}[p\!+\!q]\tilde{\bf W}_K[K_0,p])+\right.\\
&\left.\!{\mathrm D}\!(\!\sum_{p=1}^{KN_c}\!{\boldsymbol{\beta}}[p]\tilde{\bf W}_K[K_0,p\!+\!q])\!\!+\!\!{\mathrm D}(\!\sum_{p=1}^{KN_c}\!\!\tilde{\bf W}_K[K_0,p]\tilde{\bf W}_K\![K_0,\!p\!+\!q])\!\right]\\
&=\frac{1}{{|{\boldsymbol{\beta}_{\rm n}}^*|^4}}\left[{|{\boldsymbol{\beta}_{\rm c}}^*|^2}{\tilde\sigma^2}+{|{\boldsymbol{\beta}}^*|^2}{\tilde\sigma^2}+\frac{KN_c}{2}{\tilde\sigma^2}\right]\\
&\approx \frac{2{\tilde\sigma^2}}{{|{\boldsymbol{\beta}_{\rm n}}^*|^2}}+\frac{KN_c}{2{|{\boldsymbol{\beta}_{\rm n}}^*|^4}}{\tilde\sigma^2}.
\end{aligned}
\end{equation}\end{footnotesize}

\section{Proof of  proposition 2}\label{proof2}
Reconsider ${\bf s}$ and re-formulate it as
\begin{equation}
{\bf s}={\boldsymbol \phi}_m{\bf R},
\end{equation}
where ${\boldsymbol \phi}_m$ is represented by
\begin{equation}
{\boldsymbol \phi}_m=[\boldsymbol{\alpha}_1[m]\boldsymbol{\alpha}_{1}[m]\!,\!\cdots\!,\boldsymbol{\alpha}_l[m]\boldsymbol{\alpha}_{l^{'}}[m],\!\cdots\!,\!\boldsymbol{\alpha}_L[m]\boldsymbol{\alpha}_{L}[m]],
\end{equation} 
and ${\bf R}=[\boldsymbol{\rho}_{1,1}^{\rm T},\cdots,\boldsymbol{\rho}_{l,l^{'}}^{\rm T},\cdots,\boldsymbol{\rho}_{L,L}^{\rm T}]^{\rm T}$ 
Moreover, as we described before, when $KN_c$ is large enough, ${\boldsymbol\rho}_{l,l^{'}}$ is approximately a circular-shifted version of ${\boldsymbol\rho}_{1,1}$ for any $l$ and $l^{'}$, which can be formulated as
\begin{equation}
\boldsymbol{\rho}_{l,l^{'}}[q]=\boldsymbol{\rho}_{1,1}[q\oplus \frac{\Delta_{l,l^{'}}}{T_{\rm sam}}],\ q=1,\cdots,KN_c.
\end{equation}
Then, we rearrange the rows of the matrix ${\bf R}$ and the corresponding columns of ${\boldsymbol \phi}_m$ in the ascending order of $\frac{\Delta_{l,l^{'}}}{T_{\rm sam}}$, while we define the rearranged matrix as $\tilde{\bf R}$ and $\tilde{\boldsymbol \phi}_m$. 
Thus, $\boldsymbol{\rho}_{1,1}$ should satisfy the following condition
\begin{equation}\label{condition2} 
\tilde{\boldsymbol \phi}_m\tilde{\bf R}
={\bf s}.
\end{equation}

To find the specific $\boldsymbol{\rho}_{1,1}$ satisfying (\ref{condition2}), we first consider the structure of $\tilde{\bf R}$. Since each row of $\tilde{\bf R}$ is a right-circular-shifted version of $\boldsymbol{\rho}_{1,1}$, we can denote the number of right cyclic shifts corresponding to the $n$th row of $\tilde{\bf R}$ as $p_n$ for $n=1,\cdots,L^2$. Then, we define $\boldsymbol{\Lambda}_q$ as the location  indication matrix of the $q$th element of $\boldsymbol{\rho}_{1,1}$ in $\tilde{\bf R}$ and $q=1,\cdots,KN_c$. Concretely, $\boldsymbol{\Lambda}_q$ belongs to $\mathbb{C}^{L^2\times KN_c}$ and all of its $(n+q,p_n)$th elements for $n=0,\cdots,L^2-1$ are set as $1$, while the other elements are set as $0$. Furthermore, according to the definition, we have the following equation
\begin{equation}\label{lambda}
\boldsymbol{\Lambda}_{n}=\boldsymbol{\Lambda}_{n-1}{\bf J}, \ {\rm for}\ n=2,\cdots,KN_c,
\end{equation}
where ${\bf J}$ is the circulant matrix:
\begin{small}\begin{equation}\label{J}
{\bf J}=
\begin{pmatrix}
&{\bf 0}_{KN_c-1} &{\bf I}_{KN_c-1}\\
&1 &{\bf 0}_{KN_c-1}
\end{pmatrix}.
\end{equation}\end{small}

As a result, (\ref{condition2}) can be reformulated as
\begin{small}
\begin{equation}\label{matrix}
\tilde{\boldsymbol \phi}_m\tilde{\bf R}=\tilde{\boldsymbol \phi}_m\boldsymbol{\Lambda}_1\sum_{n=1}^{KN_c}\boldsymbol{\rho}_{1,1}(n){\bf J}^{n-1},
\end{equation}\end{small}utilizing  (\ref{lambda}) and (\ref{J}). Let us denote $\tilde{\boldsymbol \phi}_m\boldsymbol{\Lambda}_1$ as $\breve{\boldsymbol \phi}_m$. Then, the right-hand-side of (\ref{matrix}) can be viewed as the circular convolution between $\breve{\boldsymbol \phi}_m$ and $\boldsymbol{\rho}_{1,1}$, which is $\breve{\boldsymbol \phi}_m\circledast \boldsymbol{\rho}_{1,1}$. Furthermore, the circular convolution can be reformulated as
\begin{small}\begin{equation}
\begin{aligned}
\tilde{\boldsymbol \phi}_m\tilde{\bf R}=\breve{\boldsymbol \phi}_m\circledast \boldsymbol{\rho}_{1,1}=\boldsymbol{\rho}_{1,1}\circledast \breve{\boldsymbol \phi}_m
=\boldsymbol{\rho}_{1,1}\sum_{n=1}^{KN_c}\breve{\boldsymbol \phi}_m[n]{\bf J}^{n-1}={\bf s}.
\end{aligned}
\end{equation}\end{small}

For brevity, let us define $\breve{\bf J}=\sum_{n=1}^{KN_c}\breve{\boldsymbol \phi}_m[n]{\bf J}^{n-1}$, which is a circulant matrix. 
According to \cite{gray2006toeplitz}, a circulant matrix can be diagonalized by the discrete Fourier transform matrix and thus, $\breve{\bf J}$ can be represented as
\begin{equation}\label{diag}
\breve{\bf J}={\bf F}{\mathrm{diag}}({\breve{\boldsymbol \phi}}_{\rm F}){\bf F}^{\rm H},
\end{equation}
where ${\breve{\boldsymbol \phi}}_{\rm F}={\bf \breve{{\boldsymbol \phi}}}_m{\bf F}$.
However, due to the randomness of $\breve{\boldsymbol \phi}_m$, ${\breve{\boldsymbol \phi}}_{\rm F}$ may have zero elements, implying that we cannot tell, whether $\breve{\bf J}$ is a full-rank matrix. As a result, we discuss the value of $\boldsymbol{\rho}_{1,1}$ in terms of whether $\breve{\bf J}$ is of full rank.

1) If $\breve{\bf J}$ is of full rank, we can simply write $\boldsymbol{\rho}_{1,1}$ as
\begin{equation}\label{rhor}
\boldsymbol{\rho}_{1,1}={\bf s}\breve{\bf J}^{-1}.
\end{equation}

2) If $\breve{\bf J}$ is not full rank, we can obtain $\boldsymbol{\rho}_{1,1}$ by solving:
\begin{equation}\label{rhor4}
\boldsymbol{\rho}_{1,1}{\bf F}{\mathrm{diag}}({\breve{\boldsymbol \phi}}_{\rm F}){\bf F}^{\rm H}={\bf s}.
\end{equation}
After some simple derivations, $\boldsymbol{\rho}_{1,1}$ can be expressed as
\begin{equation}\label{rhor3}
\boldsymbol{\rho}_{1,1}={\bf s}{\bf F}{\mathrm{diag}}({\breve{\boldsymbol \phi}}_{\rm F}^{\rm Inv}){\bf F}^{\rm H},
\end{equation}
where ${\breve{\boldsymbol \phi}}_{\rm F}^{\rm Inv}$ is defined as
\begin{small}\begin{equation}\label{equ51}
\begin{cases}
{\breve{\boldsymbol \phi}}_{\rm F}^{\rm Inv}[n]{\breve{\boldsymbol \phi}}_{\rm F}[n]=1,\ {\rm for}\ {\breve{\boldsymbol \phi}}_{\rm F}[n]\neq 0,\\
{\breve{\boldsymbol \phi}}_{\rm F}^{\rm Inv}[n]=0,\ {\rm for}\ {\breve{\boldsymbol \phi}}_{\rm F}[n]= 0.
\end{cases}
\end{equation}\end{small}Moreover, the full rank $\breve{\bf J}$ also applies to (\ref{equ51}). Therefore, by defining $\breve{\bf J}^{\rm Inv}$ as 
\begin{equation}\label{newequ51}
\breve{\bf J}^{\rm Inv}={\bf F}{\mathrm{diag}}({\breve{\boldsymbol \phi}}_{\rm F}^{\rm Inv}){\bf F}^{\rm H},
\end{equation}
we can finally represent $\boldsymbol{\rho}_{1,1}$ as
\begin{equation}\label{rho_finally}
\boldsymbol{\rho}_{1,1}={\bf s}\breve{\bf J}^{\rm Inv},
\end{equation} 
regardless whether $\breve{\bf J}$ is of full rank or not.

%

\section{Proof of the proposition 3}\label{proof3}
Since the autocorrelation sequence of a discrete sequence is equivalent to the convolution of the sequence and its reversed version \cite{sundararajan2016discrete}, we can reformulate $\boldsymbol{\rho}_{1,1}$ in (\ref{rho}) as
\begin{equation}\label{rhor1}
\boldsymbol{\rho}_{1,1}=[{\boldsymbol{\bar\psi}}_{K,N_c}{\bf F}_{KN_c}]\circledast{\mathrm {Rev}}[{\boldsymbol{\bar\psi}}_{K,N_c}{\bf F}_{KN_c}],
\end{equation}where ${\mathrm {Rev}}[\cdot]$ is the operator that inverts a discrete sequence. 

By substituting (\ref{rhor1}) into (\ref{rho_finally}), we arrive at
\begin{equation}\label{rhor2}
[{\boldsymbol{\bar\psi}}_{K,N_c}{\bf F}_{KN_c}]\circledast{\mathrm {Rev}}[{\boldsymbol{\bar\psi}}_{K,N_c}{\bf F}_{KN_c}]\!=\!{\bf s}\breve{\bf J}_{\rm Inv}.
\end{equation}
Moreover, by utilizing the anti-diagonal matrix ${\bf E}\in\mathbb{C}^{{KN_c}\times{KN_c}}$ defined as
\begin{equation}\label{R}
{\bf E}= \left[
\begin{array}{ccc}
0 & \cdots & 1 \\
\vdots & 1 & \vdots \\
1 & \cdots & 0 
\end{array}
\right],
\end{equation}
we can further represent (\ref{rhor2}) as
\begin{equation}\label{41}
[{\boldsymbol{\bar\psi}}_{K,N_c}{\bf F}_{KN_c}]\circledast[{\boldsymbol{\bar\psi}}_{K,N_c}{\bf F}_{KN_c}{\bf E}]\!=\!{\bf s}\breve{\bf J}_{\rm Inv}.
\end{equation}
According to the time-domain convolution theorem \cite{orfanidis1995introduction}, (\ref{41}) can be converted into
\begin{equation}\label{equ54}
{\boldsymbol{\bar\psi}}_{K,N_c}\odot[{\boldsymbol{\bar\psi}}_{K,N_c}{\bf F}_{KN_c}{\bf E}{\bf F}_{KN_c}^{\rm H}]={\bf s}\breve{\bf J}_{\rm Inv}.
\end{equation}

Moreover, the elements of ${\bf F}_{KN_c}{\bf E}{\bf F}_{KN_c}^{\rm H}$ are all constants for a constant $K$ and we denote the matrix as $\bar{\bf E}$. Therefore, to ensure the validity of (\ref{equ54}), the following conditions should be satisfied
\begin{equation}\label{equ32}
\begin{cases}
{\boldsymbol{\bar{\psi}}}_{K,N_c}\![i]\!\sum_{j=1}^{N_c}{\boldsymbol{\bar{\psi}}}_{K,N_c}[j]\bar{\bf E}[i,j]\!=\!\boldsymbol{\rho}_{1,1}[i], {\rm for}\ i=1,\cdots,N_c,\\
\boldsymbol{\rho}_{1,1}[i]=0,\ {\rm for}\ i=N_c+1,\cdots,KN_c,
\end{cases}
\end{equation} which means that there are solutions for ${\boldsymbol{\bar{\psi}}}_{K,N_c}$ to ensure the validity of (\ref{ideal})  when $\boldsymbol{\rho}_{1,1}[i]=0$ for $i=N_c+1,\cdots,KN_c$ and meanwhile the solutions equal ${\boldsymbol{\bar{\psi}}}_{K,N_c}[i]\sum_{j=1}^{N_c}{\boldsymbol{\bar{\psi}}}_{K,N_c}[j]\bar{\bf E}[i,j]$ for $i=1,\cdots,N_c$.


\ifCLASSOPTIONcaptionsoff
  \newpage
\fi

\bibliographystyle{IEEEtran}
\bibliography{IEEEabrv,reference.bib}

\begin{IEEEbiography}[{\includegraphics[width=1in,height=1.25in,clip,keepaspectratio]{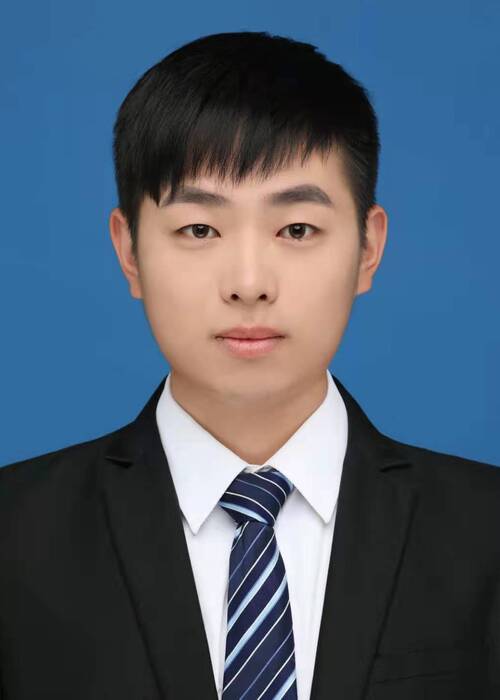}}]{Xiao-Yang Wang} received the B.S. degree in electronics and information science and technology from Beijing University of Posts and Telecommunications (BUPT), China, in 2020. He is currently pursuing the Ph.D. degree in information and communication engineering with the School of Information and Communication Engineering, BUPT, and with the Key Laboratory of Universal Wireless Communications, Ministry of Education. Since Sep. 2023, he has been a visiting Ph.D. student with the Department of Electronic and Electrical Engineering, University College London, UK. His current research interest includes integrated sensing and communications (ISAC), time/frequency synchronization, and signal processing in distributed massive MIMO systems.
\end{IEEEbiography}

\begin{IEEEbiography}[{\includegraphics[width=1in,height=1.25in,clip,keepaspectratio]{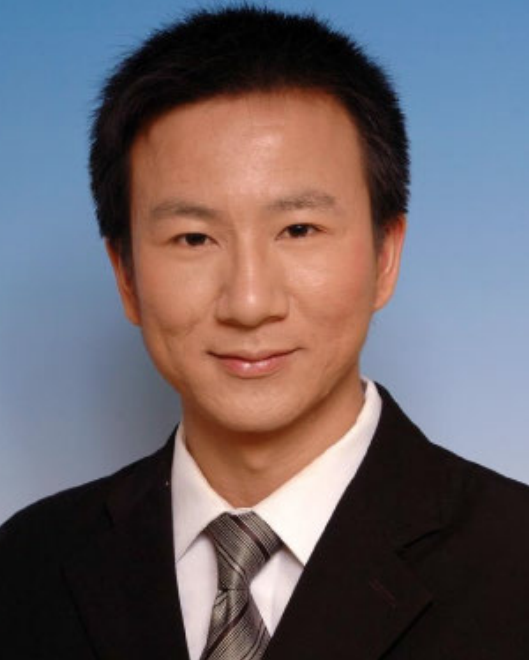}}]{Shaoshi Yang}
	(Senior Member, IEEE) 
	received the B.Eng. degree in information engineering from Beijing University of Posts and Telecommunications (BUPT), China, in 2006, and the Ph.D. degree in electronics and electrical engineering from University of Southampton, UK, in 2013. From 2008 to 2009, he was a Researcher with Intel Labs China. From 2013 to 2016, he was a Research Fellow with the School of Electronics and Computer Science, University of Southampton. From 2016 to 2018, he was a Principal Engineer with Huawei Technologies Co., Ltd., where he made significant contributions to the products, solutions and standardization of 5G, wideband IoT, and cloud gaming/VR. He was a Guest Researcher with the Isaac Newton Institute for Mathematical Sciences, University of Cambridge. He is currently a Full Professor with BUPT. His research interests include 5G/5G-A/6G, massive MIMO, mobile ad hoc networks, distributed artificial intelligence, and cloud gaming/VR. He is a standing committee member of the CCF Technical Committee on Distributed Computing and Systems. He received Dean’s Award for Early Career Research Excellence from University of Southampton in 2015, Huawei President Award for Wireless Innovations in 2018, IEEE TCGCC Best Journal Paper Award in 2019, IEEE Communications Society Best Survey Paper Award in 2020, CAI Invention and Entrepreneurship Award in 2023, CIUR Industry-University-Research Cooperation and Innovation Award in 2023, and the First Prize of Beijing Municipal Science and Technology Advancement Award in 2023. He was/is an Editor of \textit{IEEE Transactions on Communications}, \textit{IEEE Systems Journal}, \textit{IEEE Wireless Communications Letters}, and \textit{Signal Processing} (Elsevier). For more details of his research progress, please refer to https://shaoshiyang.weebly.com/.
\end{IEEEbiography}

\begin{IEEEbiography}[{\includegraphics[width=1in,height=1.25in,clip,keepaspectratio]{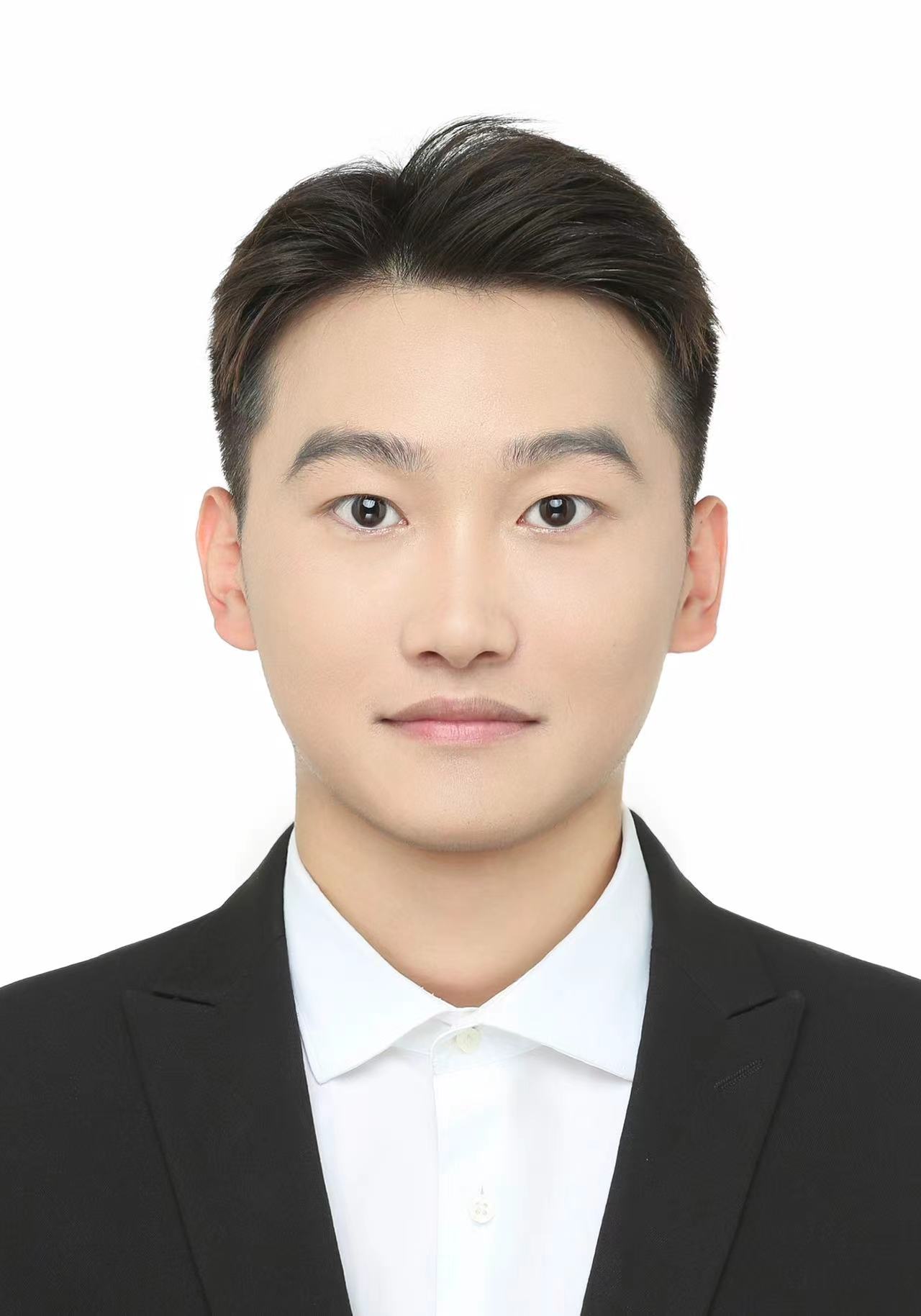}}]{Hou-Yu Zhai} received the B.S. degree in communications engineering from Beijing University of Posts and Telecommunications (BUPT), China, in 2023. He is currently pursuing the Ph.D. degree in information and communication engineering with the School of Information and Communication Engineering, BUPT, and with the Key Laboratory of Universal Wireless Communications, Ministry of Education. His current research interest includes integrated sensing and communications (ISAC), distributed beamforming, and signal processing in distributed massive MIMO systems.
\end{IEEEbiography}

\begin{IEEEbiography}[{\includegraphics[width=1in,height=1.25in,clip,keepaspectratio]{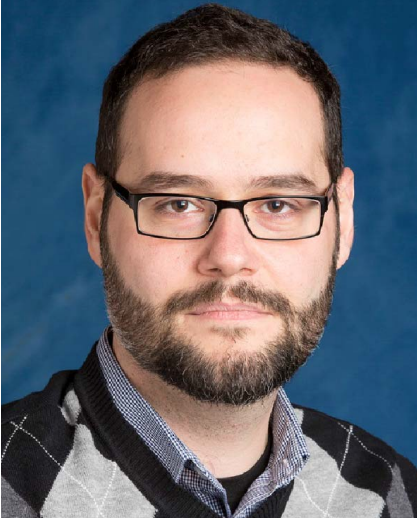}}]{Christos Masouros} (Fellow, IEEE) received the Diploma degree in electrical and computer engineering from  University of Patras, Greece, in 2004, and the M.Sc. by research and Ph.D. degrees in electrical and electronic engineering from The University of Manchester, U.K., in 2006 and 2009, respectively. In 2008, he was a Research Intern at Philips Research Labs, U.K. Between 2009 and 2010, he was a Research Associate with The University	of Manchester and a Research Fellow at Queen’s University Belfast between 2010 and 2012. In 2012, he joined University	College London as a Lecturer. He has held a Royal Academy of Engineering	Research Fellowship between 2011 and 2016. Since 2019, he has been a Full Professor of signal processing and wireless communications with the Information and Communication Engineering Research Group, Department of Electrical and Electronic Engineering, and affiliated with the Institute for Communications and Connected Systems, University College London.	His research interests lie in the field of wireless communications and signal processing with particular focus on green communications, large scale antenna systems, integrated sensing and communications (ISAC), interference mitigation techniques for MIMO, and multicarrier communications. He was the co-recipient of the 2021 IEEE SPS Young Author Best Paper Award, and the recipient of the Best Paper Awards in the IEEE GLOBECOM 2015 and IEEE WCNC 2019 conferences. He has been recognized as an Exemplary Editor for \textit{IEEE Communications Letters} and as an Exemplary Reviewer for \textit{IEEE Transactions on Communications}. He is a founding member and the Vice-Chair of the IEEE Emerging Technology Initiative on ISAC, the Vice Chair of the IEEE Special Interest Group on ISAC, and the Chair of the IEEE Special Interest Group on Energy Harvesting Communication Networks. He is an Editor of \textit{IEEE Transactions on Communications}, \textit{IEEE Transactions on Wireless Communications}, \textit{IEEE Open Journal of Signal Processing}, and the Editor-at-Large of \textit{IEEE Open Journal of the Communications Society}. He was an Associate Editor of \textit{IEEE Communications Letters} and a Guest Editor for a number of special issues on \textit{IEEE Journal of Selected Topics in Signal Processing} and \textit{IEEE Journal on Selected Areas in Communications}.
\end{IEEEbiography}

\begin{IEEEbiography}[{\includegraphics[width=1in,height=1.25in,clip,keepaspectratio]{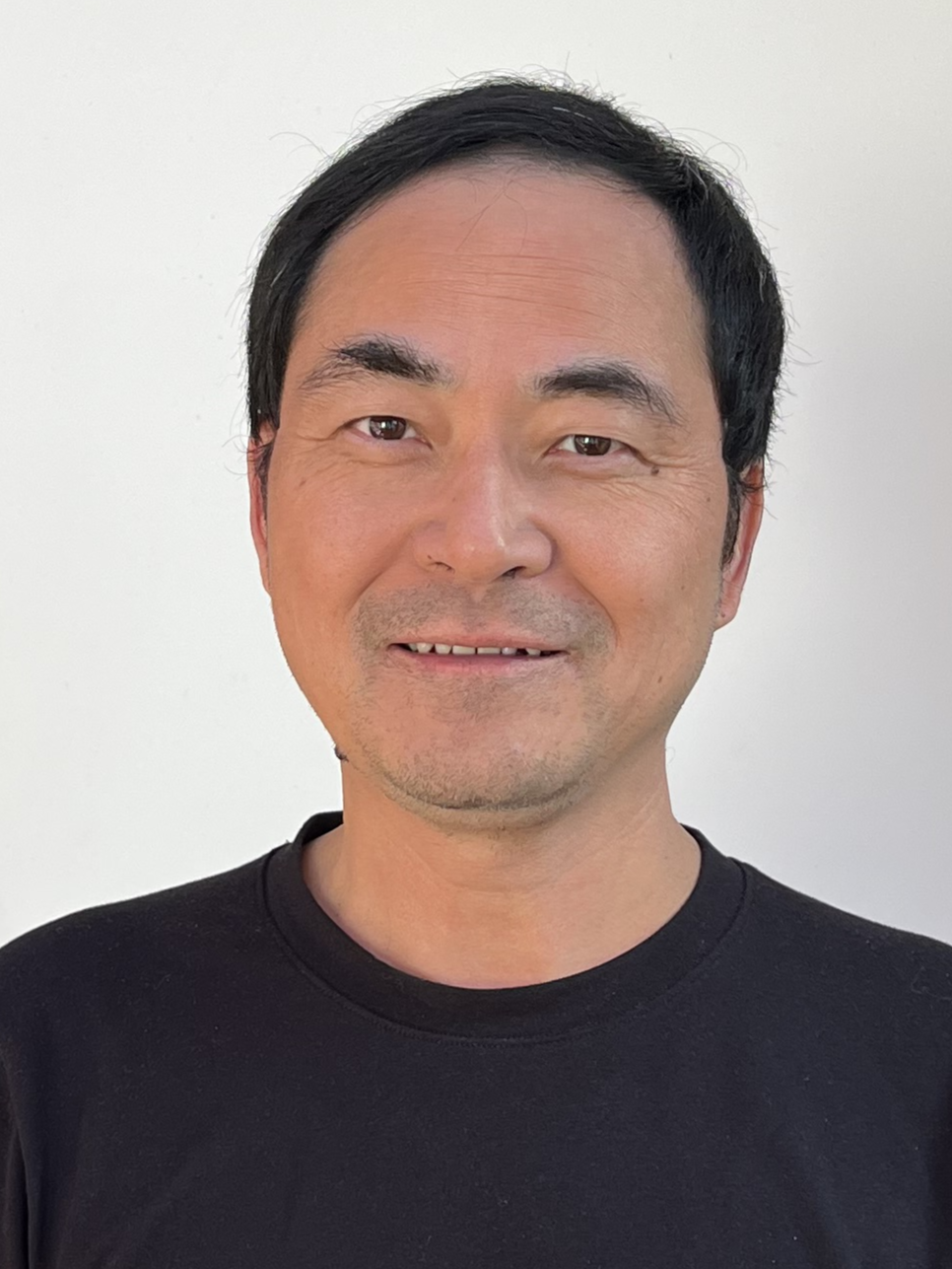}}]{J. Andrew Zhang} (Senior Member, IEEE) received the B.Sc. degree from Xi'an JiaoTong University, China, in 1996, the M.Sc. degree from Nanjing University of Posts and Telecommunications, China, in 1999, and the PhD degree from the Australian National University, Australia, in 2004.
	
Currently, Dr. Zhang is a Professor in the School of Electrical and Data Engineering, University of Technology Sydney, Australia. He was a researcher with Data61, CSIRO, Australia from 2010 to 2016, the Networked Systems, NICTA, Australia from 2004 to 2010, and ZTE Corp., Nanjing, China from 1999 to 2001.  Dr. Zhang's research interests are in the area of signal processing for wireless communications and sensing, with a focus on integrated sensing and communications. He has published more than 300 papers in leading Journals and conference proceedings, and has won 5 best paper awards for his work, including the best paper award in IEEE ICC 2013. He is a recipient of CSIRO Chairman's Medal and the Australian Engineering Innovation Award in 2012 for exceptional research achievements in multi-gigabit wireless communications.
\end{IEEEbiography}

\end{document}